\documentclass{amsart}
\usepackage{hyperref}
\usepackage{amsmath,amsthm,amscd,amssymb}
\usepackage{enumerate}

\newcommand{\arxiv}[1]{\href{http://arxiv.org/#1}{arXiv:#1}}
\newcommand*{\mailto}[1]{\href{mailto:#1}{\nolinkurl{#1}}}

\newtheorem{theorem}{Theorem}[section]
\newtheorem{lemma}[theorem]{Lemma}
\newtheorem{corollary}[theorem]{Corollary}

\numberwithin{equation}{section}

\newcommand{\C}{\mathbb{C}}
\newcommand{\R}{\mathbb{R}}

\newcommand{\E}{\mathrm{e}}
\newcommand{\I}{\mathrm{i}}

\newcommand{\siu}{\sigma^{\mathrm{u}}}
\newcommand{\sil}{\sigma^{\mathrm{l}}}
\newcommand{\sipmu}{\sigma_\pm^{\mathrm{u}}}
\newcommand{\sipml}{\sigma_\pm^{\mathrm{l}}}
\newcommand{\sipmul}{\sigma_\pm^{\mathrm{u,l}}}
\newcommand{\simpul}{\sigma_\mp^{\mathrm{u,l}}}
\newcommand{\lau}{\lambda^{\mathrm{u}}}
\newcommand{\lal}{\lambda^{\mathrm{l}}}

\newcommand{\dom}{\mathfrak{D}}

\renewcommand{\Im}{\mathop{\rm Im}}
\renewcommand{\Re}{\mathop{\rm Re}}

\newcommand{\clos}{\mathop{\rm clos}}
\newcommand{\inte}{\mathop{\rm int}}
\newcommand{\beq}{\begin{equation}}
\newcommand{\eeq}{\end{equation}}
\newcommand{\bal}{\begin{align}}
\newcommand{\eal}{\end{align}}
\newcommand{\nn}{\nonumber}

\newcommand{\ga}{\gamma}

\newcommand{\si}{\sigma}
\newcommand{\pa}{\partial}

\newcommand{\la}{\lambda}
\newcommand{\ov}{\overline}

\DeclareMathOperator{\wronsk}{\textup{\textsf{W}}}

\def\XXint#1#2#3{{\setbox0=\hbox{$#1{#2#3}{\int}$}
     \vcenter{\hbox{$#2#3$}}\kern-.5\wd0}}

\numberwithin{equation}{section}

\begin{document}

\title[On the KdV Equation with Steplike Finite-Gap Initial Data]{On the Cauchy Problem for the
Korteweg--de Vries Equation with Steplike Finite-Gap Initial Data I. Schwartz-Type Perturbations}

\author[I. Egorova]{Iryna Egorova}
\address{B.Verkin Institute for Low Temperature Physics\\
47 Lenin Avenue\\61103 Kharkiv\\Ukraine}
\email{\mailto{iraegorova@gmail.com}}

\author[K. Grunert]{Katrin Grunert}
\address{Faculty of Mathematics\\ Nordbergstrasse 15\\ 1090 Wien\\ Austria}
\email{\mailto{katrin.grunert@univie.ac.at}}
\urladdr{\url{http://www.mat.univie.ac.at/~grunert/}}

\author[G. Teschl]{Gerald Teschl}
\address{Faculty of Mathematics\\
Nordbergstrasse 15\\ 1090 Wien\\ Austria\\ and International Erwin
 Schr\"odinger
Institute for Mathematical Physics, Boltzmanngasse 9\\ 1090 Wien\\ Austria}
\email{\mailto{Gerald.Teschl@univie.ac.at}}
\urladdr{\url{http://www.mat.univie.ac.at/~gerald/}}

\thanks{Research supported by the Austrian Science Fund (FWF) under Grant
No.\ Y330.}
\thanks{Nonlinearity {\bf 22}, 1431--1457 (2009)}


\begin{abstract}
We solve the Cauchy problem for the Korteweg--de Vries equation with
initial conditions which are steplike Schwartz-type perturbations of
finite-gap potentials under the assumption that the respective spectral bands
either coincide or are disjoint.
\end{abstract}

\maketitle

\section{Introduction}

Since the seminal work of Gardner et al.\ \cite{GGKM} in 1967 the
inverse scattering
transform is one of the main tools for solving the Korteweg--de Vries
(KdV) equation
\beq\label{KdV}
q_t = -q_{xxx} + 6 q q_x
\eeq
and numerous articles have been devoted to this subject since then.
In particular,
the case when the initial condition is asymptotically close to $0$
is well understood
and we just refer to the monographs by Eckhaus and Van Harten \cite{EVH},
 Marchenko \cite{M},
  Novikov, Manakov, Pitaevskii, and Zakharov \cite{NMPZ} or Faddeev and
   Takhtajan \cite{FT}. The same is true for the
case of steplike initial conditions which are asymptotically constant
(with different constants in
different directions), where we refer to Buslaev and Fomin \cite{BF},
Cohen \cite {C}, Cohen and Kappeler \cite{CK1} and Kappeler \cite{Kap}.
In fact, even the case where the asymptotics are given by some power-like
 behaviour (including some unbounded initial conditions)
were investigated by Bondareva, Kappeler, Perry, Shubin and, Topalov
\cite{Bo}, \cite{BS}, \cite{KPST}.
On the other hand, essentially nothing is known about the Cauchy problem
for initial conditions which
are asymptotically periodic.The first to consider a periodic background
 seem to be Kuznetsov and A.V. Mikha\u\i lov, \cite{kumi}, who
informally treated the Korteweg--de Vries equation with the Weierstra{\ss}
elliptic function
as background solution. The only known results, concerning to the
existence of the solution seem to be by Ermakova \cite{Er}, \cite{Er1} and
Firsova \cite{F4} (where the evolution of the scattering
data for periodic background was given). However, both works are incomplete
from the point of view of
a rigorous application of the inverse scattering method. Surprisingly, much
more is know about the asymptotical behavior
(assuming existence) of such solutions, see for example \cite{Ba},
\cite{Bik}--\cite{Bik2}, \cite{iu}, \cite{Kh}--\cite{KhS}, \cite{N}.
Finally we mention that in the discrete case (Toda lattice) the same
problem was completely solved in \cite{EMT1} (for corresponding
long-time asymptotics see \cite{EBM}, \cite{dkkz}, \cite{km2}, \cite{kt},
\cite{kt2}, \cite{kt3}, \cite{krt2}, \cite{vdo}).

Our aim in the present paper is to provide a rigorous treatment of the
inverse scattering transform for the KdV equation in the case
of initial conditions which are steplike Schwartz-type perturbations of
finite-gap solutions. The reason which makes the
periodic case much more difficult are the poles of the Baker--Akhiezer
functions which reflect the fact that the underlying hyperelliptic
Riemann surface is no longer simply connected. In particular, we include
 a complete discussion of the problems arising from these poles.
In order to keep our presentation within reasonable limits and to be able
 to focus on the novel features of our approach, we have chosen
to limit ourselves to the case of Schwartz-type perturbations and the
additional assumption that the mutual spectral bands either coincide or
are disjoint. While this last assumption excludes the classical case of
steplike constant background, it clearly includes the case
of short range perturbations of arbitrary finite-gap solutions.
The latter being solved to the best of our knowledge for the first time
here.

More precisely, we will prove the following result

\begin{theorem}
Let $p_\pm(x,t)$ be a real-valued finite-gap solution of the KdV equation
corresponding to the initial condition $p_\pm(x)=p_\pm(x,0)$.
Suppose that the mutual spectral bands of the one-dimensional Schr\"odinger
 operators associated with $p_+$ and $p_-$ either coincide
or  are disjoint.

Let $q(x)$ be a real-valued smooth function such that (the Schwartz
class)
\beq \label{S.2}
\pm \int_0^{\pm \infty} \left|
\frac{d^n}{dx^n} \big( q(x) - p_\pm(x)\big) \right| (1+|x|^m)dx
<\infty,\quad \forall m, n \in\mathbb{N}\cup\{0\},
\eeq
then there is a unique smooth solution $q(x,t)$ of the KdV equation
corresponding to the initial condition $q(x,0)=q(x)$ and satisfying
\beq \label{S.2t}
\pm \int_0^{\pm \infty} \left| \frac{\pa^n}{\pa
x^n} \big( q(x,t) - p_\pm(x,t)\big) \right| (1+|x|^m)dx <\infty,
\quad \forall m, n \in\mathbb{N}\cup\{0\},
\eeq
for all $t\in\R$.
\end{theorem}

We will show how to remove the spectral restriction and how to handle
the case where only a finite number of moments, respectively a finite
number of derivatives, exist in a follow-up publication \cite{ET}.

\section{Some general facts on the KdV flow}

Let $q(x,t)$ be a classical solution of the KdV equation,
 that is, all partial derivatives appearing in equation \eqref{KdV}
exist and are continuous. Moreover, suppose $q(x,t)$ and $q_x(x,t)$ are
bounded with respect to $x$ for
all $t\in\R_+$.

Introduce the Lax pair \cite{Lax}
\begin{align}\label{Lop}
L_q(t) &= -\pa_x^2 + q(x,t),\\
P_q(t) &=  -4\pa_x^3 + 6q(x,t)\pa_x +3 q_x(x,t).
\end{align}
Note that $L_q(t)$ is self-adjoint on $\dom(L_q(t))=H^2(\R)$ and $P_q(t)$
is skew-adjoint
on $\dom(P_q(t))=H^3(\R)$. Moreover, the KdV equation is equivalent to the
Lax equation
\[
\pa_t L_q(t) = [P_q(t),L_q(t)]
\]
on $H^5(\R)$.

The following result follows from classical theory of ordinary
 differential equations.

\begin{lemma}
Let $c(\la,x,t)$ and $s(\la,x,t)$ be the solutions of the
differential equation $L_q(t) u = \la u$ corresponding to the initial
conditions $c(\la,0,t)=s_x(\la,0,t)=1$ and
$c_x(\la,0,t)=s(\la,0,t)=0$.

Then $c(\la,x,t)$ and $c_x(\la,x,t)$ are holomorphic with respect to
$\la\in\C$ (for fixed $x$ and $t$) and continuously differentiable
with respect to $t$ (provided $q(x,t)$ is). Similarly for
$s(\la,x,t)$ and $s_x(\la,x,t)$.
\end{lemma}

Next, note the following property

\begin{lemma}\label{lemLPKdV}
Suppose $q(x,t)$ is three times differentiable with respect to $x$ and once
with respect to $t$.
If $L_q(t) u = \la u$ holds, then
\beq
(L_q(t)-\la) (u_t - P_q(t) u) = -(q_t + q_{xxx} - 6 q q_x) u
\eeq
\end{lemma}

\begin{proof}
Suppose $L_q u = \la u$, then we have $P_q u = (2(q+2\la)\pa_x - q_x) u$
and
thus
\[
(L_q(t)-\la) P_q(t) u = (q_{xxx} - 6 q q_x) u
\]
respectively
\[
(L_q(t)-\la) u_t = - q_t u
\]
which proves the claim.
\end{proof}

\begin{corollary}[\cite{M}, corollary to Lemma 4.1.1']\label{lemMar}
Suppose $q(x,t)$ is three times differentiable with respect to $x$
and once with respect to $t$. The function $q(x,t)$ satisfies the
KdV equation \eqref{KdV} if and only if the operator
\beq\label{Aop}
\mathcal{A}_q(t) = \pa_t-2(q(x,t) +2\la)\pa_x +q_x(x,t)
\eeq
transforms solutions of equation $(L_q(t)-\la)u=0$ into solutions of
the same equation.
\end{corollary}

Furthermore, we obtain

\begin{lemma}\label{lemsysLP}
Let $q(x,t)$ be a classical solution of the KdV equation \eqref{KdV}.
The system of differential equations
\begin{align} \label{sysLP}
L_q(t) u &=\la u,\\ \label{sys1}
u_t &= P_q(t) u
\end{align}
has a unique solution $u(\la,x,t)$ for any given initial conditions
$u(\la,0,0)=a_0(\la)$ and $u_x(\la,0,0)=b_0(\la)$. It will be continuous
with respect to $\la$ if $a_0$, $b_0$ are.
\end{lemma}

\begin{proof}
Write
\[
u(\la,x,t) = a(\la,t) c(\la,x,t)+ b(\la,t) s(\la,x,t),
\]
then clearly $L_q(t) u = \la u$ holds by construction, and
Lemma~\ref{lemLPKdV} implies
\[
(L_q-\la) (u_t - P_q u) =0.
\]
Hence $u_t = P_q u$ will hold if and only if
\[
a_t c + a c_t + b_t s + b s_t = a (P_q c) + b (P _q s) = 2(2\la+q)
(a c_x+ b s_x) - q_x (a c + b s)
\]
holds together with its $x$ derivative at $x=0$, that is,
\begin{align} \nn
a_t(\la,t) &= -a(\la,t)q_x(0,t) + b(\la,t) (4\la+2q(0,t)),\\ \nn
b_t(\la,t) &= b(\la,t) q_x(0,t) + a(\la,t)
\left(2(2\la+q(0,t))(q(0,t)-\la)
 -q_{xx}(0,t)\right),\\ \nn
 a(\la,0)&= a_0(\la),\\ \label{sys7}
b(\la,0)&= b_0(\la).
\end{align}
 This is a system of ordinary differential equations for the
unknown functions $a(\la,t)$, $b(\la,t)$ and hence the claim
follows.
\end{proof}

Let $c(\la,x,t) + m_\pm(\la,t) s(\la,x,t)$ be a pair of Weyl
solutions for operator $L_q(t)$, where $m_\pm(\la,t)$ are the Weyl
$m$-functions
associated with $L_q$.

\begin{lemma}\label{lemweylLP}
The functions
\beq\label{eqphi}
u_\pm(\la,x,t) = a_\pm(\la,t) \big( c(\la,x,t) + m_\pm(\la,t) s(\la,x,t)
 \big),
\eeq
where
\beq
a_\pm(\la,t)= \exp\left(\int_0^t \Big(2\big(q(0,s)+2\la\big)m_\pm(\la,s)
 -q_x(0,s) \Big)ds \right),
\eeq
solve \eqref{sysLP}, \eqref{sys1}.
\end{lemma}

\begin{proof}
Let $u$ denote one of the Weyl solutions $u_+(\la,x,t)$ or $u_-(\la,x,t)$ and let $\bar u$ be the other one.
Then Lemma~\ref{lemLPKdV} implies that $u_t -P_q u$ is again a solution of $L_q u = \la u$. Consequently
$u_t -P_q u=\beta u + \gamma \bar u$, where $\beta=\beta(\la,t)$, $\gamma=\gamma(\la,t)$.
Since the Weyl solution decays sufficiently fast with respect to $x$ on the corresponding half-axis when $\la\in\mathbb{C}\setminus \si$, then $u_t -P_q u$ also decays on the same half-axis. Therefore, $\gamma=0$ and $u_t-P_q u =\beta u$
and the function $\hat u(\la,x,t)= \exp(-\int_0^t\beta(\la,s)ds) u(\la,x,t)$ satisfies the system \eqref{sysLP}, \eqref{sys1}.

It remains to compute $\beta(\la,t)$.
Using $u(\la,x,t)=c(\la,x,t)+m(\la,t) s(\la,x,t)$, where $m(\la,t)$ is the corresponding Weyl function, we obtain
\begin{align*}
c_t +m_t s + m s_t =& -4c_{xxx} - 4 m s_{xxx}
+ 6 q(c_x + m s_x) + 3 q_x(c+m s) + \beta(c+m s)\\
= & 4(\la c_x - q_x c - c_x q)
+ 4 m (\la s_x - q_x s - s_x q) + 6q(c_x + m s_x)\\
& + 3 q_x ( c+ms) +\beta (c+ms).
\end{align*}
For $x=0$ this equation reads $0=2(q(0,t)+2 \la) m(\la,t) - q_x(0,t) + \beta(\la,t)$.
\end{proof}

Let $\wronsk(f,g)(x)=f(x)g'(x)-f'(x)g(x)$ denote the Wronski
determinant. The next lemma is a straightforward calculation.

\begin{lemma}\label{lemW}
Let $u_1$, $u_2$ be two solutions of \eqref{sysLP}, \eqref{sys1},
then the Wronskian $\wronsk(u_1,u_2)$ does neither depend on $x$ nor
on $t$.
\end{lemma}

\section{Some general facts on finite-gap potentials}
\label{secfgp}

Since we want to study the initial value problem for the KdV equation
in the class of initial conditions which asymptotically look
like (different) finite-gap solutions, we need to recall some necessary
background
from finite-gap solutions first. For further information and for the history
 of
finite-gap solutions we refer to, for example, \cite{GH}, \cite{GRT},
\cite{M}, or \cite{NMPZ}.

Let $L_\pm(t):=L_{p_\pm}(t)$ be two one-dimensional Schr\"odinger operators
associated with two arbitrary quasi-periodic finite-gap solutions
$p_\pm(x,t)$ of the KdV equation.
We denote by
\beq\label{psin}
\psi_\pm(\la,x,t)=c_\pm(\la,x,t)+ m_\pm(\la,t)s_\pm(\la,x,t)
\eeq
the corresponding Weyl solutions of $L_\pm(t)\psi_\pm=\la\psi_\pm$,
 normalized according to
$\psi_\pm(\la,0,t)=1$ and satisfying $\psi_\pm(\la,.,t)\in
L^2((0,\pm\infty))$ for $\la\in\mathbb{C}\setminus\mathbb{R}$.

It is well-known that the spectra $\si_\pm:=\si(L_\pm(t))$ are $t$ independent
and consist of a finite number, say $r_\pm+1$, bands:
\begin{equation}\label{1.61}
\sigma_\pm = [E_0^\pm, E_1^\pm]\cup\dots\cup[E_{2j-2}^\pm,
E_{2j-1}^\pm]\cup\dots\cup[E_{2r_\pm}^\pm,\infty).
\end{equation}
Then $p_\pm$ are uniquely determined by their associated Dirichlet
divisors
\[
 \left\{(\mu_1^\pm(t),\si_1^\pm(t)), \dots,(\mu_{r_\pm}^\pm(t),
 \si_{r_\pm}^\pm(t))\right\},
\]
where $\mu_j^\pm(t) \in [ E_{2j-1}^\pm, E_{2j}^\pm]$ and $\si_j^\pm(t)
\in \{+1,-1\}$.

Let us cut the complex plane along the spectrum $\sigma_\pm$ and
denote the upper and lower sides of the cuts by $\sipmu$ and
$\sipml$. The corresponding points on these cuts will be denoted by
$\lau$ and $\lal$, respectively. In particular, this means
\[
f(\lau) := \lim_{\varepsilon\downarrow0} f(\lambda+\I\varepsilon),
\qquad f(\lal) := \lim_{\varepsilon\downarrow0}
f(\lambda-\I\varepsilon), \qquad \lambda\in\sigma_\pm.
\]
Set
\beq \label{1.0}
Y_\pm(\la)=-\prod_{j=0}^{2r_\pm} (\la-E_j^\pm),
\eeq
and introduce the functions
\begin{equation}\label{1.88}
g_\pm(\la,t)= -\frac{\prod_{j=1}^{r_\pm}(\la - \mu_j^\pm(t))}{2
Y_\pm^{1/2}(\la)},
\end{equation}
where the branch of the square root is chosen such that
\begin{equation}\label{1.8}
\frac{1}{\I} g_\pm(\lau) = \Im(g_\pm(\lau))  >0 \quad
\mbox{for}\quad \lambda\in\sigma_\pm.
\end{equation}

The functions $\psi_\pm$  admit two other well-known representations
that will be used later on. The first one is
\beq\label{1.23}
\psi_\pm(\la,x,t)= u_\pm(\la,x,t)\E^{\pm\I\theta_\pm(\la)x}
\quad\la\in\C\setminus\si_\pm
\eeq
where  $\theta_\pm(\la)$ are the
quasimoments and the functions $u_\pm(\la,x,t)$ are quasiperiodic
with respect to $x$ with the same basic frequencies as the
potentials $p_\pm(x,t)$. The quasimoments are holomorphic for
$\la\in\C\setminus\si_\pm$ and normalized according to
\beq\label{1.24}
\frac{d\theta_\pm}{d\la}>0 \quad
\mbox{for}\quad\la\in\sipmu,\qquad \theta_\pm(E_0^\pm)=0.
\eeq
This normalization implies (cf.\ \eqref{1.8})
\beq\label{1.25}
\frac{d\theta_\pm}{d\la}=\frac{\I\prod_{j=1}^{r_\pm}(\la -
\zeta_j^\pm)}{ Y_\pm^{1/2}(\la)},\qquad \zeta_j^\pm\in(E_{2j-1}^\pm,
E_{2j}^\pm),
\eeq
and therefore, the quasimoments are real-valued on
$\si_\pm$. Note, in the case where $p_\pm(x,t)\equiv0$ we have
$\theta_\pm(\la)=\sqrt{\la}$ and $u_\pm(\la,x,t)\equiv 1$.

Furthermore, the Weyl solutions possess more complicated properties,
for example, they can have poles, as we see from the other
representation. Namely, let $\mathbb{P}_\pm$ be the Riemann
surfaces, associated with the functions $Y_\pm^{1/2}(\la)$ and let
$\pi_\pm$ be parameters on these surfaces, corresponding to the
spectral parameter $\la$, where $\pi_+$ (resp. $\pi_-$) is the
parameter on the upper (resp., lower) sheet of $ \mathbb{P}_+$
(resp.\ $\mathbb{P}_-$). Then
\beq\label{1.26} \psi_\pm(\pi_\pm,x,t)
=\exp\left(\int_0^x m_\pm(\pi_\pm, y,t)dy\right),
\eeq
where $m_\pm(\pi_\pm,x,t)$ are shifted Weyl functions (cf. \cite{L}).
Note, that the Weyl function $m_+(\la,t)$ is the branch,
corresponding to values of $m_+(\pi_+,0,t)$ and
$m_-(\la,t)=m_-(\pi_-,0,t)$. Denote the divisor of poles  (the
Dirichlet divisor) of the shifted Weyl functions by
$\sum_{j=1}^{r_\pm} (\mu_j^\pm(x,t),\si_j^\pm(x,t))$. Then the
functions $\mu_j^\pm(x,t)$ satisfy the system of Dubrovin equations
(\cite[Lem.~1.37]{GH})
\begin{align}\label{1.D1}
\frac{\pa\mu_j^\pm(x,t)}{\pa x} &=-2\si_j^\pm(x,t)
Y_{\pm,j}(\mu_j^\pm(x,t),x,t),\\ \label{1.D2}
\frac{\pa\mu_j^\pm(x,t)}{\pa t} &= -4\si_j^\pm(x,t)(p_\pm(x,t)
+2\mu_j^\pm(x,t))Y_{\pm,j}(\mu_j^\pm(x,t),x,t),
\end{align}
where
\beq\label{1.D3}
Y_{\pm,j}(\la,x,t)=\frac{Y_\pm^{1/2}(\la)(\la -\mu_j^\pm(x,t))}
{G_\pm(\la,x,t)}
\eeq
and
\beq\label{1.D4}
G_\pm(\la,x,t)=\prod_{j=1}^{r_\pm}(\la - \mu_j^\pm(x,t)).
\eeq
In \eqref{1.D2} $p_\pm(x,t)$ have to be replaced by the trace formulas
\beq
p_\pm(x,t) =  \sum_{j=0}^{2r_\pm} E^\pm_j - 2\sum_{j=1}^{r_\pm}
\mu^\pm_j(x,t).
\eeq
Moreover, the following formula holds (\cite[(1.165)]{GH})
\beq\label{1.29}
m_\pm(\la,x,t)=\frac{H_\pm(\la,x,t)\pm
Y_\pm^{1/2}(\la)}{G_\pm(\la,x,t)},
\eeq
where
\beq\label{1.30}
H_\pm(\la,x,t)=\frac{1}{2}\frac{\pa}{\pa x}G_\pm(\la,x,t).
\eeq
We will also use
\beq\label{1.291}
\breve{m}_\pm(\la,x,t)=\frac{H_\pm(\la,x,t)\mp
Y_\pm^{1/2}(\la)}{G_\pm(\la,x,t)},
\eeq
to denote the other branches of the Weyl functions on the Riemann
surfaces $\mathbb{P}_\pm$, that is, $\breve{m}_\pm(\la,x,t)=
m_\pm(\pi_\pm^*,x,t)$.
In addition,
\beq
m_\pm(\la,t)-\breve{m}_\pm(\la,t)
=\frac{\pm 2 Y_\pm^{1/2}(\la)}{G_\pm(\la,0,t)}.
\eeq

\begin{lemma} \label{lemdecomp}
The following asymptotic expansion for large $\la$ is valid
\beq\label{1.31}
\psi_\pm(\la,x,t)=\exp\left(\pm\I \sqrt\la x
+\int_0^x\kappa_\pm(\la,y,t)dy\right),
\eeq
where
\beq\label{1.32}
\kappa_\pm(\la,x,t)=\sum_{k=1}^\infty\frac
{\kappa_k^\pm(x,t)}{(\pm 2\I\sqrt\la)^k},
\eeq
with coefficients defined recursively via
\beq\label{1.33}
\kappa_1^\pm(x,t)=p_\pm(x,t),\quad\kappa_{k+1}^\pm(x,t)=-\frac{\pa}
{\pa x}\kappa_k^\pm(x,t) - \sum_{m=1}^{k-1}\kappa_{k-m}^\pm(x,t)
\kappa_m^\pm(x,t).
\eeq
\end{lemma}

\begin{proof}
By \eqref{1.29} we conclude that
\[
m_\pm(\la,x,t) = \pm\I\sqrt\la + \kappa_\pm(\la,x,t),
\]
where $\kappa_\pm(\la,x,t)$ has an asymptotic expansion of the
type \eqref{1.32}.
Inserting this expansion into the Riccati equation
\beq\label{kap}
\frac{\pa} {\pa x}\kappa_\pm(\la,x,t) \pm 2\I\sqrt\la\kappa_\pm(\la,x,t)
+\kappa_\pm^2(\la,x,t) -p_\pm(x,t)=0
\eeq
and comparing coefficients shows \eqref{1.33}.
\end{proof}

As a special case of Lemma~\ref{lemweylLP} we obtain

\begin{lemma}\label{lemweyl1}
The functions
\beq\label{1.37}
\hat\psi_\pm(\la,x,t) = \E^{\alpha_\pm(\la,t)} \psi_\pm(\la,x,t),
\eeq
where
\beq\label{1.38}
\alpha_\pm(\la,t) := \int_0^t \left(2(p_\pm(0,s) + 2\la)
m_\pm(\la,s) - \frac{\pa p_\pm(0,s)}{\pa x}\right)ds,
\eeq
satisfy the system of equations
\begin{align}\label{LP1}
L_\pm(t)\hat\psi_\pm &= \la\hat\psi_\pm,\\ \label{LP2}
\frac{\pa\hat\psi_\pm}{\pa t} &= P_\pm(t)\hat\psi_\pm,
\end{align}
where $P_\pm(t):=P_{p_\pm}(t)$.
\end{lemma}

We note that (\cite[(1.148)]{GH})
\beq\label{1.38a}
\alpha_\pm(\la,t) = \frac{1}{2} \log\left(\frac{G_\pm(\la,0,t)}
{G_\pm(\la,0,0)}\right) \pm
2 Y_\pm^{1/2}(\la) \int_0^t \frac {p_\pm(0,s)+2\la}{G_\pm(\la,0,s)}ds
\eeq
and corresponding to $\breve{m}_\pm(\la,t)$ we also introduce
\begin{align}\nn
\breve{\alpha}_\pm(\la,t) & :=\int_0^t \left((2p_\pm(0,s) +
4\la)\breve{m}_\pm(\la,s) - \frac{\pa p_\pm(0,s)}{\pa x}\right)ds\\
\label{1.38new}
&=  \frac{1}{2} \log\left(\frac{G_\pm(\la,0,t)}{G_\pm(\la,0,0)}\right) \mp
2 Y_\pm^{1/2}(\la) \int_0^t \frac {p_\pm(0,s)+2\la}{G_\pm(\la,0,s)}ds.
\end{align}
Note
\beq
\overline{\alpha_\pm(\la,t)} = \breve{\alpha}_\pm(\la,t), \qquad
\la\in\si_\pm.
\eeq
In order to remove the singularities of the functions $ \psi_\pm(\la,x,t)$
we set
\beq\label{Mset}
\begin{array}{lll} M_\pm(t) &=&\{
\mu^\pm_j(t) \mid \mu^\pm_j(t) \in (E_{2j-1},E_{2j})
\text{ and } m_\pm(\la,t) \text{ has a simple pole}\},\\
\hat M_\pm(t) &=&\{ \mu^\pm_j(t) \mid \mu^\pm_j(t) \in \{E_{2j-1},
E_{2j}\} \},
\end{array}
\eeq
and introduce the functions
\begin{align} \nn
\delta_\pm(\la,t) &:= \prod_{\mu^\pm_j(t) \in M_\pm(t)}(\la-\mu^\pm_j(t)),\\
\label{S2.6}
\hat \delta_\pm(\la,t) &:= \prod_{\mu^\pm_j(t) \in M_\pm(t)}
(\la-\mu_j^\pm(t)) \prod_{\mu^\pm_j(t) \in \hat M_\pm(t)}
 \sqrt{\la - \mu^\pm_j(t)},
\end{align}
where $\prod =1$ if the index set is empty.

\begin{lemma}\label{lemalpha}
For each $t\geq 0$ and $\la\in\C\setminus \si_\pm$ the functions
$\alpha_\pm(\la,t)$ possess the properties
\begin{align}\label{alpha2}
\exp\big(\alpha_\pm(\la,t) + \breve{\alpha}_\pm(\la,t)\big) &=
\frac{G_\pm(\la,0,t)}{G_\pm(\la,0,0)},\\ \label{alpha1}
\exp\big(\alpha_\pm(\la,t)\big) &= \frac{\hat\delta_\pm(\la,t)}
{\hat\delta_\pm(\la,0)}f_\pm(\la,t),
\end{align}
where the functions $f_\pm(\la,t)$ are holomorphic in $\C \setminus \si_\pm$,
continuous up to the boundary and $f_\pm(\la,t)\neq 0$ for all $\la\in\C$.

Furthermore, let $E\in\{E_{2j-1}^\pm, E_{2j}^\pm\}$, then
\beq\label{alpha4}
\lim_{\la\to E} \left(\alpha_\pm(\la,t) -
\breve{\alpha}_\pm(\la,t)\right)=\begin{cases}
0, &\mu_j^\pm(t)\neq E, \mu_j^\pm(0)\neq E,\\
0, &\mu_j^\pm(t)= E, \mu_j^\pm(0)=E,\\
\I\pi,&\mu_j^\pm(t)=E,\mu_j^\pm(0),\neq E,\\
\I\pi,&\mu_j^\pm(t)\neq E, \mu_j^\pm(0)=E,
\end{cases}\quad \pmod{2\pi\I}.
\eeq
\end{lemma}

\begin{proof}
To shorten notations let us denote the derivative with respect to
$t$ by a dot
and the derivative with respect to $x$ by a prime. Equations
 \eqref{1.38} and \eqref{1.38a}
immediately give \eqref{alpha2} and
\beq\label{6.10}
\alpha_\pm(\la,t) - \breve{\alpha}_\pm(\la,t)=\pm
4 Y_\pm^{1/2}(\la)\int_0^t\frac {p_\pm(0,s)+2\la}{G_\pm(\la,s)}ds,
\eeq
where we have abbreviated
\[
G_\pm(\la,t):=G_\pm(\la,0,t).
\]
This function is well-defined on the set
$\C\setminus\cup_{j=1}^{r_\pm}[E_{2j-1}^\pm, E_{2j}^\pm]$, but may
have singularities inside gaps. Note, that \beq\label{alpha3}
\alpha_\pm(\la,t) -
\breve{\alpha}_\pm(\la,t)\in\R,\quad\mbox{for}\quad
\la\in\R\setminus\si_\pm. \eeq Consider the behavior of this
function in the $j$th gap. By splitting the integral $\int_0^t$ in
the definition of $\alpha_\pm(\la,t)$ (resp.\
$\breve{\alpha}_\pm(\la,t)$) into a sum of smaller integrals
$\int_{t_0}^{t_1}$ it suffices to consider the cases where
$\mu_j^\pm(s) \not\in \{E_{2j-1}^\pm, E_{2j}^\pm\}$ for
$s\in[t_0,t_1)$ or $s\in(t_0,t_1]$. We will only investigate the
first case (the other being completely analogous) and assume $t_0=0$
without loss of generality. In other words, it suffices to consider
the case where $\mu_j^\pm(0)\in (E_{2j-1}^\pm, E_{2j}^\pm)$ and the
time $t>0$ is so small, that $\si_j^\pm(s)=\si_j^\pm(0)$ for $s\leq
t$. Consequently, $\mu_j^\pm(t)\in (E_{2j-1}^\pm, E_{2j}^\pm)$ and
there exists some $\varepsilon=\varepsilon(t)$ such that
\beq\label{mu}
\mu_j^\pm(s)\in (E_{2j-1}^\pm+2\varepsilon,
E_{2j}^\pm-2\varepsilon), \quad 0\leq s\leq t.
\eeq
Consider (e.g.) the case where the point $\mu_j^\pm(s)$ moves to the right, that is
$\mu_j^\pm(0)<\mu_j^\pm(t)$. If $\la\notin
(\mu_j^\pm(0)-\varepsilon, \mu_j^\pm(t)+\varepsilon)$, then the
integral \eqref{6.10} is well-defined and by definition \eqref{1.0}
the first case of \eqref{alpha4} is fulfilled. Now let
\beq\label{mu1}
\la\in (\mu_j^\pm(0)-\varepsilon,
\mu_j^\pm(t)+\varepsilon). \eeq From equation \eqref{1.D2} we have
\beq\label{6.12}
\dot\mu_j^\pm(s)=-\sigma_j^\pm(s)\tilde
Y_{\pm,j}(\mu_j^\pm(s),s), \eeq where \beq\label{6.19}\tilde
Y_{\pm,j}(\la,s)= 4(p_\pm(s) + 2\la)Y_{\pm,j}(\la,0,s)
\eeq
and the functions $Y_{\pm,j}(\la,0,s)$ are defined by \eqref{1.D3}.
Recall that $\sigma_j^\pm(s)= \mbox{const}$. Thus
\begin{align}\nn
& \int_{0}^{t}\frac
{\pm 4(p_\pm(s) +2\la) Y^{1/2}_\pm(\la)}{G_\pm(\la,s)}=
\pm\int_{0}^{t}\frac{ \tilde Y_{\pm,j}(\la,s)}{\la - \mu_j^\pm(s)}ds\\
\label{6.18}
& \qquad = \pm\int_{0}^{t}\frac{\tilde
Y_{\pm,j}(\mu_j^\pm(s),s)}{\la - \mu_j^\pm(s)}ds \pm
\int_{0}^{t}\frac{\pa}
{\pa\la} \tilde Y_{\pm,j}(\la,s)_{\left|\la=\xi_j^\pm(s)\right.}ds,
\end{align}
where $\xi_j^\pm(s)\in(E_{2j-1}^\pm+\varepsilon, E_{2j}^\pm -
\varepsilon)$. Therefore $\frac{\pa}{\pa\la} \tilde Y_{\pm,j}(\la,s)$ is
bounded here. But
\begin{align*}
\pm\int_{0}^{t}\frac{\tilde Y_{\pm,j}(\mu_j^\pm(s),s)}{\la - \mu_j^\pm(s)} ds
&= \mp\si_j^\pm(0)\int_{0}^{t} \frac{\dot\mu_j^\pm(s)}{\la - \mu_j^\pm(s)}ds\\
&=\pm\si_j^\pm(0)\log\frac{\la -\mu_j^\pm(t)}{\la -\mu_j^\pm(0)}.
\end{align*}
Thus, in the case under consideration we have
\beq\label{6.15}
\alpha_\pm(\la,t) -
\breve{\alpha}_\pm(\la,t)=\log\frac{(\la
-\mu_j^\pm(t))^{\pm\si_j^\pm (t)}} {(\la
-\mu_j^\pm(0))^{\pm\si_j^\pm(0)}} + \tilde
f_\pm(\la,\varepsilon),
\eeq
where $\tilde f_\pm(\la,\varepsilon)$ is a smooth function, bounded by
virtue of \eqref{mu1}.
Combining this formula with \eqref{alpha2} we arrive at the following
representation:
\beq\label{6.16}
\exp\big(2\alpha_\pm(\la,t)\big)=\frac{(\la
-\mu_j^\pm(t))^{\pm\si_j^\pm(t)+1}} {(\la
-\mu_j^\pm(0))^{\pm\si_j^\pm(0)+1}}f_\pm^{(1)}(\la,t), \quad
f_\pm^{(1)}(\la,t)\neq 0,
\eeq
which is valid provided \eqref{mu} and \eqref{mu1} hold.
According to our notations $\mu_j^\pm(s)\in M_\pm(s)$
iff $\pm\si_j^\pm(s)=1$. Thus, if $\mu_j^\pm(t)\in
M_\pm(t)$ (resp.\ $\mu_j^\pm(0)\in M_\pm(0)$), then the function
$\exp(\alpha_\pm(\la,t))$ has a first order zero (resp.\ pole) at such
 a point
and does not have any other poles or zeros inside the gap $(E_{2j-1}^\pm,
 E_{2j}^\pm)$.
But if $\pm\si_j^\pm(t)=-1$ (resp.\ $\pm\si_j^\pm(0)=-1$), then the
function
$\exp(\alpha_\pm(\la,t))$ has no zero (resp.\ pole) at this point.

Now let us turn to the case $\mu_j^\pm(t)$ or $\mu_j^\pm(0)\in
\{E_{2j-1}^\pm, E_{2j}^\pm\}$.
Here we cannot use the decomposition \eqref{6.18} since
the function $\frac{\pa}{\pa\la} \tilde Y_{\pm,j}(\la,s)$ is not
bounded at the
edges of the spectrum $\si_\pm$. Suppose, that $\mu_j^\pm(0)\in
(E_{2j-1}^\pm, E_{2j}^\pm)$, the point $\mu_j^\pm(s)$ moves to the
right, and the time $t>0$ is such, that $\si_j^\pm(s)=\si_j^\pm(0)$
for $s< t$ and $\mu_j^\pm(t)=E_{2j}^\pm$.  Set
$\varepsilon<1/2(\mu_j^\pm(0) - E_{2j-1}^\pm)$ and let $\la$ be such that
\[
E_{2j-1}^\pm+\varepsilon<\la< E_{2j}^\pm+\varepsilon<E_{2j+1}^\pm.
\]
Represent the function $\tilde Y_{\pm,j}(\la,s)$, defined by \eqref{6.19},
 as
\beq\label{mu10}
\tilde Y_{\pm,j}(\la,s)=\sqrt{\la - E_{2j}^\pm}\ \breve
Y_{\pm,j}(\la,s),
\eeq
with
\beq\label{6.38}
\breve Y_{\pm,j}(\la,s)=\breve Y_{\pm,j}(\mu_j^\pm(s),s) + (\la -
\mu_j^\pm(s))\frac{\pa}{\pa\la}\breve Y_{\pm,j}(\zeta_j^\pm(s),s)
\eeq
where $\frac{\pa}{\pa\la}\breve Y_{\pm,j}$ is evidently bounded. From
\eqref{6.12} it follows
that
\[
\breve Y_{\pm,j}(\mu_j^\pm(s),s)=-\frac{\sigma_j^\pm(0)
\dot\mu_j^\pm(s)}{\sqrt{\mu_j^\pm(s) - E_{2j}^\pm}},\quad 0\leq s\leq t,
\]
and
\begin{align}\nn
\int_{0}^{t}\frac{\tilde Y_{\pm,j}(\la,s)}{\la - \mu_j^\pm(s)}ds &=
-\sigma_j^\pm(0) \sqrt{\la - E_{2j}^\pm}\left(\int_{0}^{t} \! \frac{
\dot\mu_j^\pm(s)}{\sqrt{\mu_j^\pm(s) - E_{2j}^\pm} (\la -
\mu_j^\pm(s))}ds +f_j^\pm(t,\varepsilon)\right)\\ \nn
&= -\si_j^\pm \sqrt{ E_{2j}^\pm-\la}\left(
\int_{\mu_j^\pm(0)}^{E_{2j}^\pm}\frac{d\tau} {(\la - \tau) \sqrt{
E_{2j}^\pm- \tau}}+f_j^\pm(t,\varepsilon)\right)=\\ \label{6.22}
&=\sigma_j^\pm \sqrt{ E_{2j}^\pm-\la} \int_{\sqrt{
E_{2j}^\pm-\mu_j^\pm(0)}}^0\frac{2 d y}{y^2 +\la-E_{2j}^\pm}
+O\left(\sqrt{\la - E_{2j}^\pm}\right).
\end{align}
To compute the first summand in \eqref{6.22} we will distinguish two
 cases. First let $\la\in\si_\pm$, that is, $\la>E_{2j}^\pm$.
 Then the first summand in \eqref{6.22} is equal to
\[
-2 \sigma_j^\pm(0) \I \arctan\frac{\sqrt{E_{2j}^\pm-\mu_j^\pm(0)}}{\sqrt{\la-E_{2j}^\pm}}
\to -\si_j^\pm(0) \I \pi,
\quad \mbox{as}\quad \la\to E_{2j}^\pm,\quad \la\in\si_\pm.
\]
This proves the two lower cases in \eqref{alpha4}.  Next, consider the
case when $\la\in (\mu_j^\pm(0), E_{2j}^\pm)$. Then
\[
\sigma_j^\pm(0)\sqrt{ E_{2j}^\pm-\la} \int_{\sqrt{
E_{2j}^\pm-\mu_j^\pm(0)}}^0\frac{2 d y}{y^2 +\la-E_{2j}^\pm}=
\]
\[
=\sigma_j^\pm(0)\left(-\log\frac{\sqrt{ E_{2j}^\pm-\mu_j^\pm(0)} -
\sqrt{ E_{2j}^\pm-\la}}{\sqrt{ E_{2j}^\pm-\mu_j^\pm(0)} + \sqrt{
E_{2j}^\pm-\la}}+\log(-1)\right)=
\]
\beq\label{mu3}
=-\sigma_j^\pm(0)\log\frac{\la - \mu_j^\pm(0)}{ \left(\sqrt{
E_{2j}^\pm-\mu_j^\pm(0)} + \sqrt{ E_{2j}^\pm-\la}\right)^2} +
\sigma_j^\pm(0)\I\pi.
\eeq
If $\la\to E_{2j}^\pm$, then the first
summand in \eqref{mu3} vanishes, and we arrive again at
\eqref{alpha4}. If $\la$ is in a small vicinity of $\mu_j^\pm(0)$, then
\[
\pm\int_{0}^{t}\frac{ \tilde Y_{\pm,j}(\la,s)}{\la -
\mu_j^\pm(s)}ds=\mp\si_j^\pm(0)\log (\la - \mu_j^\pm(0)) + O(1),
\]
that confirm \eqref{alpha1} for the case under consideration.
\end{proof}

\section{Scattering theory}

First we collect some facts from scattering theory for
Schr\"odinger operators with step-like finite-gap potentials
(cf.\ \cite{BET}).
To shorten notations we omit the dependence on $t$ throughout this section.

Let $L_\pm$ be two Schr\"odinger operators with real-valued
finite-gap potentials $p_\pm(x)$, corresponding to the spectra
\eqref{1.61} and the Dirichlet divisors
$\sum_{j=1}^{r^\pm}(\mu_j^\pm,\sigma_j^\pm)$, where
$\mu_j^\pm\in[E_{2j-1}^\pm,E_{2j}^\pm]$ and $\sigma_j^\pm\in\{-1,
1\}$.

Let $q(x)$ be a real-valued smooth function satisfying condition
\eqref{S.2}.  The case $m=2$ and $n=0$ was rigorously studied in
\cite{BET}. In this section we point out the necessary modifications
for the Schwartz case. Let
\begin{equation}\label{S.12}
L_q :=- \frac{d^2}{dx^2} +q(x),\quad x\in \R,
\end{equation}
be the ``perturbed" operator with a potential $q(x)$, satisfying
\eqref{S.2}. The spectrum of $L_q$ consists of a purely absolutely
continuous
part $\sigma:=\sigma_+\cup\sigma_-$ plus a finite number of eigenvalues
situated
in the gaps, $\sigma_d\subset\R\setminus\sigma$. We will use the
notation $\inte(\sigma_\pm)$ for the interior of the spectrum, that
is, $\inte(\sigma_\pm):=\sigma_\pm\setminus\pa\sigma_\pm$. The set
$\sigma^{(2)}:=\sigma_+\cap\sigma_-$ is the spectrum of multiplicity two,
and $\sigma_+^{(1)}\cup\sigma_-^{(1)}$ with $\sigma_\pm^{(1)}=
\clos(\sigma_\pm\setminus\sigma_\mp)$
is the spectrum multiplicity one.

The Jost solutions of the equation
\begin{equation}\label{S.4}
\left(-\frac{d^2}{dx^2}+q(x)\right)y(x)= \la y(x),\quad \la\in \C,
\end{equation}
that are asymptotically close to the Weyl solutions of the background
operators as $x\to\pm\infty$,
can be represented with the help of the transformation operators as
\begin{equation}\label{S2.2}
\phi_\pm(\la,x) =\psi_\pm(\la,x)\pm\int_{x}^{\pm\infty}
 K_\pm(x,y)\psi_\pm(\la,y) dy,
\end{equation}
where $K_\pm(x,y)$ are real-valued functions, that satisfy the
 integral equations
\begin{align}\nn
 K_\pm(x,y)& = -2\int_{\frac{x+y}{2}}^{\pm\infty} \left(q(s) -
  p_\pm(s)\right)D_\pm(x,s,s,y)ds \\ \label{A.1}
& \mp2 \int_{x}^{\pm\infty}ds\int_{y\pm x\mp s}^{y\pm s\mp x}D_\pm(x,s,r,y)
K_\pm(s,r) \left(q(s) - p_\pm(s)\right)dr, \quad \pm y>\pm x,
\end{align}
where
\begin{equation}\label{A.2}
 D_\pm(x,y,r,s)=\mp\frac{1}{4}\sum_{E\in\partial\sigma_\pm}
 \frac{f_\pm(E,x,y)f_\pm(E,r,s)}{\frac{d}{d\la} Y_\pm(E)},
\end{equation}
with
\begin{equation}\label{A.3}
f_\pm(E,x,y)=\lim_{\la\to E}\left(\prod_{j=1}^{r_\pm}(\la-\mu_j^\pm)
\right)\psi_\pm(\la,x)\breve{\psi}_\pm(\la,y).
\end{equation}
In particular,
\begin{equation}\label{A.5}
 K_\pm(x,x)=\pm\frac{1}{2}\int_x^{\pm\infty} (q(s)-p_\pm(s))ds.
\end{equation}
Since
 \[
\frac{\pa^{n+l}}{\pa x^l\pa y^n}f_\pm(E,x,y)\in L^\infty(\mathbb{R}
\times\mathbb{R}),
\]
condition \eqref{S.2} and the method of successive approximations imply
smoothness of the kernels for the transformation operators and the
 following estimate
\beq\label{S2.3}
\left|\frac{\pa^{n+l}}{\pa x^n\pa y^l}K_\pm(x,y)\right|<\frac{C_\pm(n,l,m)}
{|x+y|^m},
\quad x,y\to\pm\infty, \quad m,n,l\in\mathbb{N}\cup \{0\},
\eeq
where $C_\pm(n,l,m)$ are positive constants (cf \cite{BET}).

Representation \eqref{S2.2} shows, that the Jost solutions inherit all
singularities of the background  Weyl
$m$-functions $m_\pm(\la)$. Hence we set (recall \eqref{S2.6})
\begin{equation}\label{S2.12}
\tilde\phi_\pm(\la,x)=\delta_\pm(\la) \phi_\pm(\la,x)
\end{equation}
such that the functions $\tilde\phi_\pm(\la,x)$ have no poles in the
 interior of
the gaps of the spectrum $\si$. Let
\[
\sigma_d=\{\lambda_1,\dots,\lambda_p\}\subset\R\setminus\sigma
\]
be the set of eigenvalues of the operator $L_q$. For every eigenvalue we
introduce the corresponding norming constants
\begin{equation} \label{S2.14}
\left(\gamma_k^\pm\right)^{-2}=\int_{\R} \tilde\phi_\pm^2(\lambda_k,x) dx.
\end{equation}
Furthermore, introduce the scattering relations
\begin{equation}\label{S2.16}
T_\mp(\lambda) \phi_\pm(\lambda,x) =\overline{\phi_\mp(\lambda,x)} +
R_\mp(\lambda)\phi_\mp(\lambda,x), \quad\lambda\in\simpul,
\end{equation}
where the transmission and reflection coefficients are defined as
usual,
\begin{equation}\label{2.17}
T_\pm(\lambda):= \frac{\wronsk(\overline{\phi_\pm(\lambda)},
\phi_\pm(\lambda))}{\wronsk(\phi_\mp(\lambda),
\phi_\pm(\lambda))},\qquad R_\pm(\lambda):= -
\frac{\wronsk(\phi_\mp(\lambda),\overline{\phi_\pm(\lambda)})}
{\wronsk(\phi_\mp(\lambda), \phi_\pm(\lambda))}, \quad\lambda\in
\sipmul.
\end{equation}

\begin{lemma}\label{lem2.3} Suppose \eqref{4.2}. Then
the scattering data
\begin{align}\nn
{\mathcal S}: = \Big\{ & R_+(\lambda),\;T_+(\lambda),\;
\lambda\in\sigma_+^{\mathrm{u,l}}; \; R_-(\lambda),\;T_-(\lambda),\;
\lambda\in\sigma_-^{\mathrm{u,l}};\\\label{S4.6} &
\lambda_1,\dots,\lambda_p\in\R\setminus \sigma,\;
\gamma_1^\pm,\dots,\gamma_p^\pm\in\R_+\Big\}
\end{align}have the following properties:
\begin{enumerate}[\bf I.]
\item
\begin{enumerate}[\bf(a)]
\item
$T_\pm(\lau) =\overline{T_\pm(\lal)}$ for $\lambda\in\sigma_\pm$.\\
$R_\pm(\lau) =\overline{R_\pm(\lal)}$ for $\lambda\in\sigma_\pm$.
\item
$\dfrac{T_\pm(\lambda)}{\overline{T_\pm(\lambda)}}= R_\pm(\lambda)$
for $\lambda\in\sigma_\pm^{(1)}$.
\item
$1 - |R_\pm(\lambda)|^2 = \dfrac{g_\pm(\lambda)}{g_\mp(\lambda)}
|T_\pm(\lambda)|^2$ for $\lambda\in\sigma^{(2)}$ with $g_\pm(\la)$
from \eqref{1.88}.
\item
$\overline{R_\pm(\lambda)}T_\pm(\lambda) +
R_\mp(\lambda)\overline{T_\pm(\lambda)}=0$ for
$\lambda\in\sigma^{(2)}$.
\item
$T_\pm(\lambda) = 1 + O\Big(\frac{1}{\sqrt\la}\Big)$ for
$\lambda\to\infty$.
\item
$R_\pm(\lambda) =
O\Big(\frac{1}{\left(\sqrt{\la}\right)^{n+1}}\Big)$ for
$\lambda\to\infty$ and for all $n\in\mathbb{N}$.

\end{enumerate}
\item
The functions $T_\pm(\lambda)$ can be extended as meromorphic
functions into the domain $\C \setminus \sigma$ and satisfy
\begin{equation}\label{S2.18}
\frac{1}{T_+(\la) g_+(\la)} = \frac{1}{T_-(\la) g_-(\la)}=:-W(\la),
\end{equation}
where the function $W(\la)$ possesses the following properties:
\begin{enumerate}[\bf(a)]
\item
The function $\tilde W(\la)=\delta_+(\la)\delta_-(\la) W(\la)$,
where $\delta_\pm(\la)$ is defined by \eqref{S2.6}, is holomorphic
in the domain $\C\setminus\sigma$, with simple zeros at the points
$\lambda_k$, where
\begin{equation}\label{S2.11}
\biggl(\frac{d\tilde W}{d \la}(\lambda_k)\biggr)^2
=(\gamma_k^+\gamma_k^-)^{-2}.
\end{equation}
In addition, it satisfies
\begin{equation}\label{S2.9}
\overline{\tilde W(\lau)}=\tilde W(\lal), \quad
\lambda\in\sigma\quad \text{and}\quad \tilde W(\lambda)\in\R
\quad \text{for} \quad \lambda\in\R\setminus \sigma.
\end{equation}
\item
The function $\hat W(\la) = \hat\delta_+(\la) \hat\delta_-(\la)
W(\la)$, where $\hat \delta_\pm(\la)$ is defined by \eqref{S2.6}, is
continuous on the set $\C\setminus\sigma$ up to the boundary
$\siu\cup\sil$. Moreover, the  function $\hat W(\la)$ is infinitely
many
 times
differentiable with respect to $\la$ on the set
$\left(\siu\cup\sil\right)\setminus
\pa\si$ and continuously differentiable with respect to the local
variable $\sqrt{\la - E}$ for $E\in\pa\si$. It can have zeros on the set
$\pa\sigma$ and does not
vanish at the other points of the set $\sigma$.  If $\hat W(E)=0$ as
$E\in\pa\sigma$, then
$\hat W(\la) =  \sqrt{\la -E} (C(E)+o(1))$, $C(E)\ne 0$.
\end{enumerate}
\item
\begin{enumerate}[\bf(a)]
\item
The reflection coefficients $R_\pm(\lambda)$ are continuously
differentiable infinitely many time
functions on the sets $\inte(\sipmul)$.
\item
If $E\in\pa\sigma$ and $\hat W(E)\neq 0$ then
the functions $R_\pm(\lambda)$ are also continuous at $E$. Moreover,
in this case
\begin{equation}\label{P.1}
R_\pm(E)=
\begin{cases}
-1 &\text{for } E\notin\hat M_\pm,\\
1 &\text{for } E\in\hat M_\pm.
\end{cases}
\end{equation}
\end{enumerate}
\end{enumerate}
\end{lemma}

\begin{proof}
For the case $m=2$ and $n=0$ this lemma was proven in \cite{BET}.
In particular,
except for the differentiability properties of the scattering data and item
{\bf I.(f)} everything follows from Lemma~3.3 in \cite{BET}.

Differentiability of $\hat W(\la)$ and $R_\pm(\la)$ is a direct
consequence of differentiability of the
Jost solutions. In fact, since $\frac{\pa^l\psi_\pm(\la,y)}{\pa\la^l}=
O(|y|^l)$ for $\la\in\inte
\si_\pm$ as $y\to\pm\infty$, equations \eqref{S2.2}, \eqref{S2.3}, and
\eqref{S.2} imply, that $\phi_\pm(\la,x)$ are continuously
differentiable infinitely many times with respect to $\la\in\inte \si_\pm$
since $\psi_\pm(\la,x)$ are.
Moreover, note, that at the points $E_j^\pm$ these solutions are
continuously differentiable with respect to the local parameter
$\sqrt{\la - E_j^\pm}$ since this holds for $\psi_\pm(\la,x)$. Furthermore,
since $\Im\theta_\pm(\la)>0$ for
$\la\in\R\setminus\si_\pm$, we infer that $\psi_\pm(\la,y)$ are
exponentially decaying together with all derivatives as
$y\to\pm\infty$ if $\la\in\R\setminus\si_\pm$.

It remains to show {\bf I.(f)}. To this end, represent the Jost solutions
 in the form
\beq\label{Jost1}
\phi_\pm(\la,x)=\psi_\pm(\la,x)\exp\left(-\int_x^{\pm\infty}
\tilde\kappa_\pm(\la,y) dy \right),
\eeq
where
\beq\label{kappa}
\tilde\kappa_\pm(\la,x)=\sum_{k=1}^
\infty\frac{\tilde\kappa_k^\pm(x)}{(\pm 2\I\sqrt\la)^k}.
\eeq
To derive a differential equation for $\tilde\kappa_\pm(\la,x)$ we
substitute \eqref{Jost1} into \eqref{S.4} and use \eqref{1.31} and
\eqref{kap}. This yields the differential equations
\beq\label{difkap}
\frac{\pa} {\pa x}\tilde\kappa_\pm(\la,x)
+\tilde\kappa_\pm^2(\la,x) \pm 2(\I\sqrt\la
+ \kappa_\pm(\la,x))\tilde\kappa_\pm(\la,x) +p_\pm(x)-q(x)=0,
\eeq
from which we obtain the recurrence formulas
\beq\label{S1.33}
\tilde\kappa_1^\pm(x)=q(x) -p_\pm(x),\quad\tilde\kappa_{k+1}^\pm(x)=-
\frac{\pa}{\pa x}
\tilde\kappa_k^\pm(x) - \sum_{m=1}^{k-1}\tilde\kappa_{k-m}^\pm(x)
(\tilde\kappa_m^\pm(x)
+2\kappa_m^\pm(x)).
\eeq
Using \eqref{2.17} we now derive an asymptotic formula for $R_+(\la)$
(for $R_-$ the considerations are
analogous). By \eqref{Jost1} and \eqref{kappa}
\beq\label{denom}
\wronsk(\phi_-(\lambda), \phi_+(\lambda))= \phi_-(\la,0)\phi_+(\la,0)
\left(2\I\sqrt\la
+O\left(\frac{1}{\sqrt\la}\right)\right)=2\I\sqrt\la (1 + o(1))
\eeq
and
\beq\label{nom}
\wronsk(\phi_-(\lambda),\overline{\phi_+(\lambda)})=
\phi_-(\lambda,0) \overline{\phi_+(\lambda,0)}
\left(\ov{y_+(\la,0)}-y_-(\la,0)\right),
\eeq
where we have set $y_\pm(\la,x):=\tilde\kappa_\pm(\la,x) +
 \kappa_\pm(\la,x)$.
Equations \eqref{kap} and \eqref{difkap} imply
\beq\label{dify}
\frac{\pa} {\pa x}y_\pm(\la,x)\pm 2\I\sqrt\la y_\pm(\la,x)
+y_\pm^2(\la,x) -q(x)=0.
\eeq
Therefore, the functions $\tilde y_+(\la,x):=\ov{y_+(\la,x)}$
and $\tilde y_-(\la,x):=y_-(\la,x)$
satisfy one and the same equation. Moreover, $\kappa_1^\pm(x)+
\tilde\kappa_1^\pm(x)=q(x)$.
Hence, since $q(x)$ is smooth, the functions $\tilde y_\pm$
admit asymptotic expansions
\[
\tilde y_\pm(\la,x)=\sum_{k=1}^\infty\frac{\tilde y_k^\pm(x)}
{(-2\I\sqrt\la)^k},
\]
where $\tilde y_k^+(x)$ and $\tilde y_k^-(x)$ satisfy the
same recurrence equations
\beq\label{recur}
\tilde y_1^\pm(x)=q(x),\quad \tilde y_{k+1}^\pm(x)
= - \frac{\pa}{\pa x}\tilde y_k^\pm(x) - \sum_{l=1}^{k-1}\tilde
y_{k-l}^\pm(x) \tilde y_l^\pm(x).
\eeq
Therefore,
\[
\ov{y_+(\la,0)}-y_-(\la,0) =O(\la^{-n/2})
\]
for $\lambda\to\infty$ and for all $n\in\mathbb{N}$ and the same is
true for $R_+(\la)$ by \eqref{denom} and \eqref{nom}.
\end{proof}

To complete the characterization of scattering data $\mathcal{S}$,
consider the associated Gelfand-Levitan-Marchenko (GLM) equations.

\begin{lemma}\label{lem4.2}
The kernels $K_\pm(x,y)$ of the transformation operators satisfy
the Gelfand-Levitan-Marchenko equations
\begin{equation}\label{ME}
K_\pm(x,y) + F_\pm(x,y) \pm \int_x^{\pm\infty} K_\pm(x,s)
F_\pm(s,y)d s =0, \quad \pm y>\pm x,
\end{equation}
where \footnote{Here we have used
the notation
$\oint_{\sigma_\pm}f(\lambda)d\la := \int_{\sipmu} f(\lambda)d\la -
\int_{\sipml} f(\lambda)d\la$.}
\begin{align}\label{4.2}
F_\pm(x,y) &= \frac{1}{2\pi\I}\oint_{\sigma_\pm}
R_\pm(\lambda) \psi_\pm(\lambda,x) \psi_\pm(\lambda,y)
g_\pm(\lambda)d\la + \\ \nn &\quad + \frac{1}{2\pi
\I}\int_{\sigma_\mp^{(1),\mathrm{u}}} |T_\mp(\lambda)|^2
\psi_\pm(\lambda,x) \psi_\pm(\lambda,y)g_\mp(\lambda)d\la\\ \nn
&\quad + \sum_{k=1}^p (\gamma_k^\pm)^2 \tilde\psi_\pm(\lambda_k,x)
\tilde\psi_\pm(\lambda_k,y).
\end{align}

\begin{enumerate}[\bf I.]
\addtocounter{enumi}{3}
\item
The functions $F_\pm(x,y)$ are  differentiable infinitely many
times with respect to both variables and satisfy
\beq\label{4.3}
\left|\frac{\pa^{l+n}}{\pa x^l\pa y^n}F_\pm(x,y)\right|\leq
\frac{C_\pm(m,n,l)}{|x+y|^m}\quad\mbox{as}\ \ x,y\to\pm\infty,\quad
m,l,n=0,1,2,\dots
\eeq
\end{enumerate}
\end{lemma}

\begin{proof}
Formulas \eqref{ME} and \eqref{4.2} are obtained in \cite{BET},
estimate \eqref{4.3} follows directly from
\eqref{ME} and \eqref{S2.3}.
\end{proof}

Properties \textbf{I--IV} from above are characteristic for the
scattering data $\mathcal S$, that is

\begin{theorem}[characterization, \cite{BET}]\label{theor1}
Properties \emph{\textbf{I--IV}} are necessary and sufficient for a
set $\mathcal{S}$ to be the set of scattering data for operator $L$
with a potential $q(x)$ from the class \eqref{S.2}.
\end{theorem}

In addition, we will now describe a procedure of solving of the
inverse scattering problem.

Let $L_\pm$ be two one-dimensional finite-gap Schr\"odinger
operators associated with the potentials $p_\pm(x)$. Let $\mathcal{S}$
be given scattering data \eqref{S4.6} satisfying \textbf{I--IV} and define
corresponding kernels $F_\pm(x,y)$ via \eqref{4.2}.
As it shown in \cite{BET},
 condition \emph{\textbf{IV}} the GLM equations \eqref{ME} have
unique smooth real-valued solutions $K_\pm(x,y)$, satisfying  estimate
of type \eqref{S2.3}, possibly with some other constants $C_\pm$, than
in \eqref{4.3}.
In particular,
\begin{equation}\label{5.101}
\pm\int_0^{\pm\infty} (1+|x|^m)\left|\frac{d^n}{dx^n}
K_\pm(x,x)\right|
dx<\infty, \qquad \forall m,n\in\mathbb{N}.
\end{equation}
Now introduce the functions
\begin{equation}\label{5.1}
q_\pm(x) =\mp 2\frac{d}{dx}K_\pm(x,x) + p_\pm(x),\quad x\in\R
\end{equation}
and note that the estimate \eqref{5.101} reads
\begin{equation}\label{5.2}
\pm \int_0^{\pm \infty}|\frac{d^n}{dx^n}(q_\pm(x) - p_\pm(x))|
(1+|x|^m) d x <\infty ,\quad \forall n,m\in\mathbb{N}\cup \{0\}.
\end{equation}
Moreover, define functions $\phi_\pm(\la,x)$ by formula \eqref{S2.2},
 where
$K_\pm(x,y)$ are the solutions of \eqref{ME}. Then these functions
solve the equations
\begin{equation}\label{5.3}
\left(-\frac{d^2}{d x^2} + q_\pm(x)\right) \phi_\pm(\la,x) =
\la\phi_\pm(\la,x).
\end{equation}
The only remaining difficulty is to show that in fact $q_-(x)=q_+(x)$:

\begin{theorem}[\cite{BET}]\label{theor2}
Let the scattering data ${\mathcal S}$, defined as in \eqref{S4.6},
satisfy the properties \emph{\textbf{I--IV}}. Then the
functions $q_\pm(x)$, defined by \eqref{5.1} coincide,
$q_-(x)\equiv q_+(x)=:q(x)$. Moreover, the data ${\mathcal S}$
are the scattering data for the Schr\"odinger operator with
potential $q(x)$ from the class \eqref{S.2}.
\end{theorem}

\section{The inverse scattering transform}

As our next step we show how to use the solution of the inverse scattering
problem found in the previous section to give a formal scheme for solving
the initial-value problem for the KdV equation with initial data from the
class \eqref{S.2}.

Suppose first that our initial-value problem has a solution $q(x,t)$
satisfying \eqref{S.2t} for each $t>0$. Then all considerations from
the previous section apply to the operator $L_q(t)$ if we consider
$t$ as an additional parameter. In particular, there are
time-dependent transformation operators with kernels $K_\pm(x,y,t)$
satisfying the estimates
\begin{equation}\label{S2.3t}
\left|\frac{\pa^{l+n}}{\pa x^l\pa y^n}K_\pm (x,y,t)\right|\leq
\frac{C_\pm(m,n,l,t)}{|x+y|^m},\quad x,y\to\pm\infty,\quad l,n,m=0,1,2,
\dots.
\end{equation}
and
\beq\label{dert}
\left\vert \frac{\partial^{n+l+1}}{\partial x^n \partial y^l\partial t}
K_\pm(x,y,t)\right\vert \leq
\frac{C_\pm(m,n,l,t)}{|x+y|^m},\quad x,y\to\pm\infty,\quad l,n,m=0,1,2,
\dots.
\eeq
These estimates follows from the fact that the kernels
$D_\pm(x,y,s,r,t)$ of the time-dependent
equations \eqref{A.1} are smooth with respect to all variables,
 and each partial derivative is uniformly
bounded with respect to $x,y,s,r,t\in\R$. Consequently, the Jost
solutions
\beq\label{Jost t}
\phi_\pm(\la,x,t) =\psi_\pm(\la,x,t)\pm\int_{x}^{\pm\infty} K_\pm(x,y,t)
\psi_\pm(\la,y,t) dy,
\eeq
are also differentiable with respect to $t$ and satisfy
\begin{align} \label {jost t1}
\frac{\pa}{\pa t}\phi_\pm(\la,x,t) &= \frac{\pa}{\pa t}\psi_\pm(\la,x,t)
 ( 1 + o(1)) \qquad\text{as }
x\to\pm\infty,\\ \label {jost t2}
\frac{\pa^n}{\pa x^n}\phi_\pm(\la,x,t) &= \frac{\pa^n}{\pa x^n}\psi_\pm
(\la,x,t) ( 1 + o(1))
\qquad\text{as } x\to\pm\infty.
\end{align}
By Lemma~\ref{lemLPKdV} we know that the functions
$P_q(t)\phi_\pm(\la,x,t)$ solves the equation $L_q(t)u=\la u$.
Asymptotics \eqref{jost t1} and \eqref{jost t2} show, that
\[
P_q(t)\phi_\pm(\la,x,t)=\beta_\pm(\la,t)\phi_\pm(\la,x,t),
\]
where $\beta_\pm(\la,t)$ is the same factor as in
$P_\pm(t)\psi_\pm(\la,x,t)=\beta_\pm(\la,t)\psi_\pm(\la,x,t)$. From
Lemma~\ref{lemweylLP} we obtain then
\begin{lemma}\label{jostevol}
Let $\alpha_\pm(\la,t)$ be defined by \eqref{1.38} and let $q(x,t)$
be
 a solution of the KdV
equation satisfying \eqref{S.2}. Then the functions
\beq\label{2.37}
\hat\phi_\pm(\la,x,t)= \E^{\alpha_\pm(\la,t)} \phi_\pm(\la,x,t)
\eeq
solve the system \eqref{sysLP}, \eqref{sys1}.
\end{lemma}

Before we proceed further we note that equation \eqref{alpha1} implies

\begin{corollary}\label{hatphi}
The function $\hat\phi_\pm(\la,x,t)$,
defined by formula \eqref{2.37}, have simple poles on the set
$M_\pm(0)$, square root singularities on the set $\hat M_\pm(0)$,
and no other singularities.
\end{corollary}

Next, consider the time-dependent scattering relations
\beq\label{S2.16t}
T_\mp(\lambda,t) \phi_\pm(\lambda,x,t) = \overline{\phi_\mp(\lambda,x,t)}
 +
R_\mp(\lambda,t)\phi_\mp(\lambda,x,t), \quad\lambda\in\simpul.
\eeq
Then, using the previous lemma in combination with Lemma~\ref{lemW}
 to evaluate
\eqref{2.17} we infer

\begin{lemma}\label{evolution}
Let $q(x,t)$ be a solution of the KdV equation satisfying \eqref{S.2}.
Then $\la_k(t)=\la_k(0)\equiv \la_k;$
\begin{align}\label{refl}
R_\pm(\la,t) &= R_\pm(\la,0)\E^{\alpha_\pm(\la,t)
-\breve{\alpha}_\pm(\la,t)}, \quad \la\in\si_\pm, \\ \label{trans}
T_\mp(\la,t) &= T_\mp(\la,0)\E^{\alpha_\pm(\la,t)
-\breve{\alpha}_\mp(\la,t)},\quad\la\in\C,\\ \label{norm}
\left(\gamma_k^\pm(t)\right)^2 &= \left(\gamma_k^\pm(0)\right)^2 \,
\frac{\delta_\pm^2(\la_k,0)}{\delta_\pm^2(\la_k,t)}\,
\E^{2\alpha_\pm(\la_k,t)},
\end{align}
where $\alpha_\pm(\la,t)$, $\breve{\alpha}_\pm(\la,t)$,
 $\delta_\pm(\la,t)$
are defined in \eqref{1.38}, \eqref{1.38new}, \eqref{S2.6}, respectively.
\end{lemma}

\begin{proof}
First of all set $\hat{W}(\la,t)= \hat\delta_+(\la,t) \hat\delta_-(\la,t)
W(\la,t)$ (recall \eqref{S2.18}). Then, since $\wronsk(\hat\phi_-(\la,t), \hat\phi_+(\la,t))$
does not depend on $t$ by Lemma~\ref{lemW}, it follows from \eqref{S2.18}  and
\eqref{alpha1} that
\beq\label{zerosW}
f(\la,t) \hat{W}(\la,t)= \hat{W}(\la,0),\quad f(\la,t)= f_-(\la,t) f_+(\la,t)\neq 0.
\eeq
This implies, that the discrete spectrum of the operator $L(t)$,
which is the set of zeros of the function $\hat{W}(\la,t)$ on the
set $\R\setminus\si$, does not depend on $t$.

Similarly, if we replace the functions $\phi_\pm$ by $\hat\phi_\pm$
in all Wronskians of formulas \eqref{2.17}, the result will be a
constant with respect to $t$. Together with \eqref{2.37} it implies
 \eqref{refl} and \eqref{trans}. To obtain \eqref{norm} we
set $\check{\phi}(\la,x,t) = \delta_\pm(\la,0)
\hat\phi_\pm(\la,x,t)$ (which is continuous near $\la_k$) and
compute
\begin{align*}
\frac{d}{dt} \int_\R \check\phi_\pm(\la_k,x,t)^2 dx &= 2 \int_\R
\check\phi_\pm(\la_k,x,t) \pa_t \check\phi_\pm(\la_k,x,t) dx\\
& = \int_\R \check\phi_\pm(\la_k,x,t) P_q(t) \check\phi_\pm(\la_k,x,t)
 dx =0,
\end{align*}
since $P_q$ is skew-adjoint and $ \check\phi_\pm(\la_k,x,t)$ is real-valued.
Note that interchanging differentiation and integration is permissible
by the dominated convergence
theorem (recall that the quasimoments $\theta_\pm(\la)$ are independent
of $t$). Thus, \eqref{S2.12} and \eqref{S2.14}
imply
\[
\frac{d}{dt} \frac{\delta_\pm(\la_k,0)\ \E^{\alpha_\pm(\la_k,t)}}
{\delta_\pm(\la_k,t)\ \gamma_k^\pm(t)}  =0,
\]
which finishes the proof.
\end{proof}

Hence the solution $q(x,t)$ can be computed from the time-dependent
scattering data as follows. Construct one of the
functions $F_+(x,y,t)$ or $F_-(x,y,t)$ via
\begin{align}\label{6.2}
F_\pm(x,y,t) =& \frac{1}{2\pi\I}\oint_{\si_\pm}
R_\pm(\la,t) \psi_\pm(\la,x,t) \psi_\pm(\la,y,t) g_\pm(\la,t)d\la +
\\ \nn & {} +\,\frac{1}{2\pi\I}\int_{\si_\mp^{(1),\mathrm{u}}}
 |T_\mp(\la,t)|^2
\psi_\pm(\la,x,t) \psi_\pm(\la,y,t)g_\mp(\la,t)d\la \\ \nn & {} +
\sum_{k=1}^p (\ga_k^\pm(t))^2 \tilde\psi_\pm(\la_k,x,t)
\tilde\psi_\pm(\la_k,y,t).
\end{align}
Solve the corresponding GLM equation
\beq\label{ME1}
K_\pm(x,y,t) +
F_\pm(x,y,t) \pm \int_x^{\pm\infty} K_\pm(x,s,t) F_\pm(s,y,t) ds =0,
\quad \pm y>\pm x,
\eeq
and obtain the solution by
\begin{equation}\label{5.111}
q(x,t) =\mp2\ \frac{d}{dx}K_\pm(x,x,t) + p_\pm(x,t),\quad
x\in\R.
\end{equation}
Theorem \ref{theor2} guarantees, that both formulas give one and the
same solution.

Up to now we have assumed that $q(x,t)$ is a solution the KdV equation
satisfying
\eqref{S.2}. Now we can get rid of this assumption. We will proceed as
follows.
Suppose the initial condition $q(x)$ satisfies \eqref{S.2} with some
finite-gap
potential $p_\pm(x)$. Consider the corresponding scattering data
$\mathcal{S}=\mathcal{S}(0)$
which obey conditions \textbf{I}--\textbf{IV}. Let $p_\pm(x,t)$
 be the finite-gap solution of
the KdV equation with initial condition $p_\pm(x)$ and let $m_\pm(\la,t)$,
$\psi_\pm(\la,x,t)$, and $\alpha_\pm(\la,t)$ be the corresponding
quantities as in
Section~\ref{secfgp}.

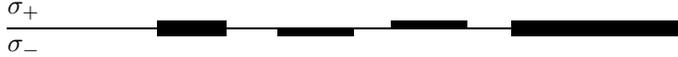
\begin{figure}[ht]
\begin{picture}(11,1.2)
\put(1,0.2){$\sigma_-$}
\put(1,0.7){$\sigma_+$}

\put(1,0.5){\line(1,0){8}}
\put(3,0.4){\rule{9mm}{2mm}}
\put(4.6,0.4){\rule{10mm}{1mm}}
\put(6.1,0.5){\rule{10mm}{1mm}}
\put(7.7,0.5){\rule{23mm}{1mm}}
\put(7.7,0.4){\rule{23mm}{1mm}}
\end{picture}
\caption{Typical mutual locations of $\sigma_-$ and $\sigma_+$.}
\label{figsi}
\end{figure}

Introduce the set of scattering data $\mathcal S(t)$, where
$R_\pm(\la,t)$, $T_\pm(\la,t)$ and $\gamma_k^\pm(t)$ are defined by
formulas \eqref{refl}--\eqref{norm}. In the next section we prove,
that these data satisfies conditions \textbf{I}--\textbf{III}, and the
functions $F_\pm(x,y,t)$, defined via \eqref{6.2}, satisfy \textbf{IV}
 under the
assumption that the respective bands of the spectra $\si_\pm$ either
 coincide or otherwise
do not intersect at all, that is
\beq\label{assump}
\sigma^{(2)}\cap\si_\pm^{(1)}=\emptyset\quad \text{and}\quad
\si_+^{(1)}\cap\si_-^{(1)}=\emptyset.
\eeq
The typical situation is depicted in Figure~\ref{figsi}.

Then Theorem 5.3 from \cite{BET} ensures the unique
solvability for each of the GLM equations \eqref{ME1}with the solutions
$K_\pm(x,y,t)$ that satisfy the estimate of type
\eqref{S2.3t}. Moreover, since $F_\pm(x,y,t)$ are differentiable
with respect to $t$ with  \eqref{4.3}  valid for this derivative,
then \eqref{ME} implies
 \eqref{dert}.
Consequently, the function $q(x,t)$, defined by formula
\eqref{5.111}, has a continuous derivative with respect to $t$ and
satisfies \eqref{S.2t} and
\beq\label{Deriv t}
\pm \int_0^{\pm\infty}
\left| \frac{\pa}{\pa t} \big( q(x,t) - p_\pm(x,t)\big)
\right| (1+|x|^m)dx <\infty.
\eeq
Moreover, the functions
$\phi_\pm(\la,x,t)$, defined via \eqref{Jost t}, solve equation
\eqref{sysLP} with $q(x,t)$, defined by \eqref{5.111}. To prove,
that this $q(x,t)$ solves the KdV equation, we will apply
Corollary~\ref{lemMar} as follows.

Since $\phi_+(\la,x,t)$ and $\phi_-(\la,x,t)$ are independent for
all $\la\in\C$ but a finite number of values, it is sufficient to
check that both functions $(\mathcal{A}_q \phi_\pm)(\la,x,t)$ solve
\eqref{sysLP}, where $\mathcal{A}_q$ is defined by \eqref{Aop} with
$q(x,t)$ from \eqref{5.111}. But due to \eqref{Jost t} and the
estimates \eqref{S2.3t}, \eqref{dert} we have \eqref{jost t1} and
\eqref{jost t2}. This implies one should show that
\beq\label{justification}
(\mathcal{A}_q \phi_\pm)(\la,x,t)
=\beta_\pm(\la,t) \phi_\pm(\la,x,t),
\eeq
for some $\beta_\pm(\la,t)$. Letting $x\to\pm\infty$ in \eqref{justification}
and comparing with
\beq\label{Apm}
(\mathcal{A}_{p_\pm}\psi_\pm)(\la,x,t)=-\frac{\pa
\alpha_\pm(\la,t)}{\pa t} \psi_\pm(\la,x,t)
\eeq
(which is evident form Lemma \ref{lemweyl1}),
gives \beq\label{just3} \beta_\pm(\la,t)= -\frac{\pa
\alpha_\pm(\la,t)}{\pa t} = -2(p_\pm(0,t)+2\la)m_\pm(\la,t) +
\frac{\pa p_\pm(0,t)}{\pa x}. \eeq Finally, as already pointed out
before, \eqref{justification} is equivalent to the KdV equation for
$q(x,t)$ by Corollary~\ref{lemMar}. Equality \eqref{justification}
will be proved in the next section.

\section{Justification of the inverse scattering transform}

Our first task is to check, that if $\mathcal{S}(0)$ satisfies
{\bf I}--{\bf III},
then the time-dependent scattering data $\mathcal{S}(t)$, defined by
\eqref{refl}--\eqref{norm} satisfy the same conditions (with
$g_\pm(\la)=g_\pm(\la,t)$). Properties {\bf I}, {\bf (a)--(f)} are
straightforward to
check. Using
\beq\label{gevol}
g_\pm(\la,t)=g_\pm(\la,0)\E^{\alpha_\pm(\la,t) +\breve{\alpha}_\pm(\la,t)},
\eeq
which follows from \eqref{1.88} and \eqref{alpha2}, we see that $W(\la,t)$
defined as in \eqref{S2.18} satisfies
\beq \label{6.29}
W(\la,t)= W(\la,0) \E^{-\alpha_-(\la,t) - \alpha_+(\la,t)}.
\eeq
Hence Lemma~\ref{lemalpha} implies that properties {\bf II}, {\bf (a)}
 and {\bf (b)}
hold.

Property {\bf III}, {\bf (a)} is evident, and property {\bf III},
{\bf (b)} follows from \eqref{alpha4}. In summary,

\begin{lemma}\label{propI-III}
Let the set $\mathcal{S}(0)$ satisfy
properties {\bf I}--{\bf III} and let the set $\mathcal{S}(t)$ be defined
by
\eqref{refl}--\eqref{norm}. Then the set $\mathcal{S}(t)$ satisfies
{\bf I}--{\bf III} with $g_\pm(\la,t)$ defined by \eqref{gevol}.
\end{lemma}

Now substitute formulas \eqref{refl}--\eqref{norm}, \eqref{1.37},
\eqref{alpha2}, and \eqref{gevol} into \eqref{6.2}, then we obtain the
following representation for the kernels of GLM equations
\begin{align}\label{6.2hat}
F_\pm(x,y,t) =& \frac{1}{2\pi\I}\oint_{\si_\pm}
R_\pm(\la,0)\,\hat\psi_\pm(\la,x,t) \hat\psi_\pm(\la,y,t)
g_\pm(\la,0)d\la\\ \nn
& {} +\,\frac{1}{2\pi\I}\int_{\si_\mp^{(1),\mathrm{u}}}
 |T_\mp(\la,0)|^2
\hat\psi_\pm(\la,x,t)\hat \psi_\pm(\la,y,t)g_\mp(\la,0)d\la \\ \nn &
{} + \sum_{k=1}^p (\ga_k^\pm(0))^2 \tilde\psi_\pm(\la_k,x,t)
\tilde\psi_\pm(\la_k,y,t),
\end{align}
where the functions
\beq\label{hattilde}
\tilde\psi_\pm(\la,x,t):= \delta_\pm(\la,0)\hat\psi_\pm(\la,x,t)
\eeq
are well-defined (bounded, continuous) for $\la\in\C\setminus\si_\pm$.
Recall that the functions $\hat\psi_\pm(\la,x,t)$ inherit all singularities
from the functions $\psi_\pm(\la,x,0)$, that is, they have simple poles
 on the set
$M_\pm(0)$, square-root singularities on the set $\hat M_\pm(0)$,
and no other singularities. Therefore, formula \eqref{6.2hat}
consists of three well-defined summands, the singularities of the
integrands are integrable (cf.\ \cite[Sect.~5]{BET}), and it remains to
verify {\bf IV}.

Due to our assumption \eqref{assump} the second and third summands in
\eqref{6.2hat} (or \eqref{6.2})
satisfies {\bf IV} for all $m$ and $n$, and hence we only need to
investigate the first summand in \eqref{6.2}.
To this end, we use \eqref{1.23}--\eqref{1.25} to obtain the
representation
\begin{align}\nn
F_{\pm,R}(x,y,t) &:= 2\Re \int_{\si_\pm^{\mathrm{u}}}
R_\pm(\la,t) \psi_\pm(\la,x,t)
\psi_\pm(\la,y,t) \frac{g_\pm(\la,t)}{2\pi\I}d\la\\ \label{Fc}
&=\Re\int_0^\infty \E^{\pm\I(x+y)\theta_\pm}
\rho_\pm(\theta_\pm,x,y,t) d\theta_\pm,
\end{align}
where
\begin{align}\label{defrho}
\rho_\pm(\theta_\pm,x,y,t) &:=\frac{1}{2\pi} \Psi_\pm(\theta_\pm,x,y,t)
\E^{\alpha_\pm(\la,t) - \breve{\alpha}_\pm(\la,t)} R_\pm(\la,0),\\
\label{Psi}
\Psi_\pm(\theta_\pm,x,y,t) &:= u_\pm(\la,x,t) u_\pm(\la,y,t)
\prod_{j=1}^{r_\pm}\frac{\la -\mu_j^\pm(t)}{\la-\zeta_j^\pm},
\end{align}
and $\la=\la(\theta_\pm)$. We will integrate \eqref{Fc} by parts $m$
times for arbitrary $m$. Since the integrand is not continuous
for $\theta_\pm\in[0,\infty)$, we regard this integral as
\beq\label{sumFc} F_{\pm,R}(x,y,t)= \Re \sum_{k=0}^{r_\pm}
\int_{\theta_\pm(E_{2k}^\pm)}^ {\theta_\pm(E_{2k+1}^\pm)}
\E^{\pm\I(x+y)\theta} \rho_\pm(\theta,x,y,t) d\theta, \eeq where we
set $E^\pm_{2 r_\pm+1}=+\infty$ for notational convenience. Then the
boundary terms during integration by parts will be
\beq\label{outint} \Re\,\lim_{\la\to
E}\frac{e^{\pm\I\theta_\pm(E)(x+y)}\frac{\pa^s\rho_\pm(\la(\theta_\pm),
x,y,t)}{\pa\theta_\pm^s}}{\left(\I(x+y)\right)^{s+1}}, \quad
s=0,1,\dots, \; E\in\pa\si_\pm, \eeq and we will prove that they
vanish for all s=0,1,....

\begin{lemma} \label{lemestim3} Let $E\in\pa\si_\pm$.
The following limits exists  for all $s=0,1,...$ and take real or
pure imaginary values: \beq\label{reflest} \lim_{\la\to
E,\,\la\in\si_\pm}\frac{d^s}{d\theta_\pm^s}\,
R_\pm(\la(\theta_\pm),0)\in\I^s\,\R,  \eeq \beq\label{bigpsi}
e^{\pm\I\,\theta_\pm(E)\,(x+y)} \lim_{\la\to
E}\frac{\pa^s}{\pa\theta_\pm^s}\, \Psi_\pm(\theta_\pm,x,y,t)\,\in
\I^s\R,  \eeq \beq\label{alphest} \lim_{\la\to
E}\frac{\pa^s}{\pa\theta_\pm^s}\,\exp\{ \alpha_\pm(\la,t) -
\check\alpha_\pm(\la,t)\}\,\in \I^s\R. \eeq
\end{lemma}

\begin{proof}
The proof is the same for $+$ and $-$ cases, we will give it for $+$
case and omit the sign $+$ in notations, except of notation for
spectrum $\si_+$.

Let $\varepsilon$ be a positive value smaller than the
minimal length of all bands in $\si_+$ and abbreviate
\[
\mathcal{O}(E)=(E-\varepsilon, E+\varepsilon)\cap\si_+.
\]
Let
\[
\mathcal{F}(E) =C^\infty(\mathcal{O}(E),\R)
\]
be the class of all functions $f(\la)$ which are smooth and real-valued
on $\mathcal{O}(E)$
and let
\[
\mathcal{G}(E) = \{ f_1(\la)+\I\frac{d\la}{d\theta}\,f_2(\la) \,|\,
f_1,f_2\in\mathcal{F}(E) \}.
\]
From \eqref{1.25} we see that $\frac{d\la}{d\theta}$ is a real-valued
and bounded
function on the set $\si_+$ and $\frac{d\la}{d\theta}(E)=0$. This function
 is smooth
with respect to $\theta$ on the set $\mathcal{O}(E)$. From
\eqref{1.24} we conclude, that
\beq\label{lasq}
\frac{d^2\la}{d\theta^2}=\frac{d}{d\la}
\left(\frac{\I\,Y^{1/2}(\la)}{\prod (\la - \zeta_j)}\right)\,\frac
{\I\,Y^{1/2}(\la)}{\prod (\la - \zeta_j)}\in\mathcal{F}(E) \text{
and } \left(\frac{d\la}{d\theta}\right)^2\in\mathcal{F}(E).
\eeq
In particular, the last two formulas imply that $\mathcal{G}(E)$ is an
algebra. Moreover, from \eqref{lasq} it follows, that
\beq\label{dertheta}
\frac{d^{2k}\la}{d\theta^{2k}}(E)\in\R,\quad
\frac{d^{2k+1}\la}{d\theta^{2k+1}}(E)=0.
\eeq
Now let
\beq\label{struct}
g(\la)= f_1(\la)+\I\frac{d\la}{d\theta}\,f_2(\la)\in\mathcal{G}(E),
\eeq
then \eqref{lasq} shows that
\beq\label{struct5}
\frac{d g(\la)}{d\theta}=\I\left(\frac{df_2}{d\la}
\left(\frac{d\la}{d\theta}\right)^2 + f_2\frac{d^2\la}{d\theta^2} -
\I\frac{d f_1}{d\la}\frac{d\la}{d\theta}\right) \in\I\mathcal{G}(E).
\eeq
Hence \eqref{struct} and \eqref{struct5} imply
\beq\label{difg}
\frac{d^s g}{d\theta^s}(E)\in \I^s\R,\quad s=0,1,\dots,
\eeq where
the values are to be understood as limits at $E$ from within the
spectrum. In particular, for any $f(\la) \in\mathcal{F}(E)$,
\beq\label{diff} \frac{d^{2k}f}{d\theta^{2k}}(E)\in\R,\quad
\frac{d^{2k+1}f}{d\theta^{2k+1}}(E)=0,\quad k=0,1,\dots. \eeq The
idea of the proof of \eqref{reflest} and \eqref{bigpsi} is to write
$R(\la,0)$ and
\beq\label{hatbigpsi}
\hat\Psi(\theta,x,y,t):= \psi(\la,x,t) \psi(\la,y,t)
\prod_{j=1}^{r}\frac{\la -\mu_j(t)}{\la- \zeta_j}
\eeq
in the form \eqref{struct}. We start with
$\hat\Psi(\theta)$ (where $x,y,t$ play the role of parameters). From
\eqref{1.30}, \eqref{1.D1}, \eqref{1.D3}, and \eqref{1.D4} we see,
 that the function
$\frac{H(\la,0,t)}{G(\la,0,t)}$ is a holomorphic function in a
vicinity of $E$ even if $\mu_j(t)=E$. Thus,
\beq\label{V}
\frac{H(\la,0,t)}{G(\la,0,t)}\in\mathcal{F}(E).
\eeq
Since
$\zeta_j\in(E_{2j-1}, E_{2j})$, then $\prod (\la -
\zeta_j)^{-1}\in\mathcal{F}.$ Also $s(\la,x,t)$,
$c(\la,x,t)\in\mathcal{F}(E)$. Using in \eqref{hatbigpsi} the
representations \eqref{psin}, \eqref{1.29}, and \eqref{1.30} we
conclude that the function $\hat\Psi(\theta,x,y,t)$ admits
a representation of the type \eqref{struct}. Therefore
\beq\label{psilim}
\lim_{\la\to E}\frac{\pa^s}{\pa\theta^s}
\hat\Psi(\theta,x,y,t) \in \I^s\R,\quad s=0,1,\dots.
\eeq
Note that in this formula it is in fact irrelevant from what side the
limit is taken.

Now consider the function $\Psi(\la,x,y,t)$ defined by formula
\eqref{Psi}. As is known (cf.\cite{BBEIM}, \cite{GH}) for each $t$
and $\la$ this function is a quasiperiodic bounded function with
respect to $x$ and $y$. Therefore, if its derivatives with respect
to the quasimomentum variable exist, then they will be bounded with
respect to $x$ and $y$. Taking into account \eqref{psilim} we obtain
\[
\lim_{\la\to E}\frac{\pa^s}{\pa\theta^s}
\Psi(\theta,x,y,t) =U_s(E,x,y,t) \E^{-\I\theta(E) (x+y)},
\]
where $U_s(E,x,y,t)\in \I^s\R$, $s=0,1,\dots$, are functions which are
bounded with respect to $x, y\in\R$ for each $t$. This
proves \eqref{Psi}. Note that $\E^{-\I\theta(E) (x+y)}$ has modulus one,
but it is in general not real-valued.

To prove \eqref{reflest} we will distinguish the resonant and
nonresonant cases. We start with nonresonant case $\hat W(E,t)\neq 0$
(cf.\ {\bf II, (b)} and note that by \eqref{zerosW} this is independent
of $t$).

Suppose, that $E\in\pa\si_+\cap\pa\si^{(2)}$ is a left edge of the
spectrum $\si$, that is,
\beq\label{Esov}
E=E_{2j}^+=E_{2k}^-.
\eeq
Consider the reflection coefficient $R_+(\la,0)$, defined by formula
\eqref{2.17} and let $\theta:=\theta_+$. Suppose, that $\mu_j^+(0)\neq
E$, $\mu_k^-(0)\neq E$. Then from \eqref{psin},
\eqref{V}, \eqref{S2.2}, \eqref{S2.3},\eqref{S.2}, and \eqref{1.25} we
see, that the Jost solution $\phi_+(\la,x)$ plus its derivative
$\frac{\pa}{\pa x} \phi_+(\la,x)$ is in $\mathcal{G}(E)$. Moreover,
by \eqref{1.25} and \eqref{Esov}
\[
\frac{d\theta_+}{d\theta_-}=\frac{d\theta_+}{d\la}\frac{d\la} {d\theta_-}
\in \frac{\sqrt{(\la -E_{2k}^-)(\la -E_{2k+1}^-)}}{\sqrt{(\la -E_{2j}^-)
(\la
-E_{2j+1}^-)}} \mathcal{F}(E) = \mathcal{F}(E)).
\]
Therefore, the same is true for $\phi_-(\la,x)$ and hence we also have
\[
\wronsk(\phi_-(\lambda), \phi_+(\lambda)), \:
\wronsk(\phi_-(\lambda),\overline{\phi_+(\lambda)}) \in \mathcal{G}(E)
\]
Since $\wronsk(\phi_-,\,\phi_+)(E)\neq 0$ we conclude $R_+(\la,0) \in
\mathcal{G}(E)$
and \eqref{reflest} is proven in this case.

If $\mu_j^+(0)\neq E$ but $\mu_k^-(0)= E$ we replace $\phi_-(\la,x)$
by
\[
\phi_-^{(1)}(\la,x):= \I\frac{d\la}{d\theta}\,\phi_-(\la,x)
\]
which is in $\mathcal{G}(E)$ and proceed as before (observe that the
 extra factor cancels in
the definition of $R_+(\la,0)$). The cases $\mu_j^+(0)= E$,
$\mu_k^-(0)\neq E$ and $\mu_j^+(0)=\mu_k^-(0)= E$ can be handled
similarly.

In the nonresonant case, when $E\in\pa\si_+^{(1)}\cap\pa\si$ the
consideration are even simpler, because in this case
(cf.\ \eqref{S2.12}) $\tilde\phi_-(\la,x)\in \mathcal{F}(E)$. We assume
$\mu_j^+(0)\neq E$, if $\mu_j^+(0)= E$ one only needs to replace
$\phi_+(\la)$
by $\phi_-^{(1)}(\la)$ as pointed out before. Thus
\beq\label{struct1}
R_+(\la,0)=\frac{f_1(\la) + \I\frac{d\la}{d\theta}f_2(\la)}
{f_3(\la)+\I\frac{d\la}
{d\theta}f_4(\la)},\quad \mbox{ where } f_i(\la)\in\mathcal{F}(E),
\,i=1,2,3,4.
\eeq
This finishes the proof of formula \eqref{reflest} in the nonresonant
case, because in this case we have $f_3(E)\neq 0$ and, therefore
$R_+(\la,0)\in\mathcal{G}(E)$.

In the resonance case we have $\hat W(E)=0$ but $\frac{d\hat
W}{d\theta}(E)\neq 0$ (cf. {\bf II, (b)}). Hence we have
\eqref{struct1} with $f_1(E)=f_3(E)=0$ and $f_4(E)\neq 0$. Let us
show that the derivative of the right-hand side of \eqref{struct1}
satisfies \beq\label{struct3} \frac{d}{d\theta}\frac{f_1(\la) +
\I\frac{d\la}{d\theta}f_2(\la)} {f_3(\la)+\I\frac{d\la}
{d\theta}f_4(\la)} \in \I\mathcal{G}(E). \eeq

Namely, denote by dot the
derivative with respect to $\theta$ and by prime - with respect to
$\la$. Then
\[
\frac{d}{d\theta}\frac{g_1(\la) + \I\dot\la
g_2(\la)}{g_3(\la)+\I\dot\la g_4(\la)}=\I\left(\ddot\la(g_2g_3 -
g_4g_1) + (\dot\la)^2(g_1^\prime g_4 - g_3^\prime g_2+ g_2^\prime
g_3 - g_4^\prime g_1)+\right.
\]
\[
\left.+\I\dot\la\left(g_3^\prime g_1 -g_1^\prime g_3 +
(\dot\la)^2(g_2^\prime g_4 - g_4^\prime
g_2)\right)\right)\left(-(\dot\la)^2\,g_4^2 +g_3^2 +
\I\dot\la(2g_4g_3)\right)^{-1}.
\]
Functions $g_1, g_3$ and $(\dot\la)^2$ have zeros of the first order
with respect to $\la$ at the point $E$ and $g_4(E)\ddot\la(E)\neq
0$. It means, that we can divide nominator and denominator in the
r.h.s. of the last equality by $(\dot\la)^2$ and using \eqref{lasq}
we arrive at \eqref{struct3}. The last one implies \eqref{reflest}
for $s\ge 1$. To prove the remaining case $s=0$ we have to check
that $R_+(E,0)\in\R$ in the resonance case. Since the nominator and
denominator in \eqref{struct1} vanishes,
\[
\lim_{\la\to E}R_+(\la,0)=\lim_{\la\to E}
\frac{(f_1^\prime+\I f_2)\dot\la +\I\ddot \la f_2}
{(f_3^\prime+\I f_4)\dot\la +\I\ddot \la f_4}=\frac{f_2(E)}{f_4(E)}\in\R.
\]
this completes the proof of \eqref{reflest}.

To prove \eqref{alphest} we use the same approach. Again the prove will be
done for the $+$ case. From \eqref{alpha4} it follows, that
\[
\lim_{\la\to E} \exp\big(\alpha_+(\la,t) - \overline{\alpha_+}(\la,t)
\big)\in\R,
\]
therefore it suffices to show that for
\[
h(\la):=\left(\alpha_+(\la,t) -
\overline{\alpha_+}(\la,t)\right)
\]
the derivative $\dot h(\la)=\frac{d h}{d\theta}$ satisfies
\beq\label{al8}
\dot h(\la)=\I f(\la), \quad f(\la)\in\mathcal{F}(E).
\eeq
To simplify notations, we will omit sign $+$ until the end of
this lemma.

Suppose first, that \beq\label{al10}\mu_j(t)\neq E=E_{2j},\quad
\mu_j(0)\neq E\eeq Let $0<t_1<...<t_N<t$ be the set of points, where
$\mu_j(t_k)=E$. Choose $\delta>0$ so small, that
\[
\mu_j(E\pm\delta)>\max\{\mu_j(0), \mu_j(t),(E_{2j-1} + E)/2\}.
\]
Denote
\[
\Delta=[0,t]\setminus\cup_{k=1}^N (t_k-\delta, t_k+\delta).
\]
Let $\la>E$ be a point in the spectrum, close to $E$. Then for
$s\in\Delta$ $|\mu_j(s) - \la|>const(E)>0$ we have (see \eqref{6.10})
\beq\label{al1}
4 Y^{1/2}(\la)\int_\Delta\frac {p_\pm(0,s)
+2\la}{G_\pm(\la,s)}ds =\I\dot\la f_1(\la),\quad
f_1\in\mathcal{F}(E).
\eeq
On the remaining set we use the
representations \eqref{mu10} and \eqref{6.38}. Proceeding as in
\eqref{6.22} we obtain
\[
4 Y^{1/2}(\la)\int_{t_k-\delta}^{t_k}\frac {p_+(0,s)
+2\la}{G_+(\la,s)}ds = -\si_j\I\left(\arctan
\frac{\sqrt{E-\mu_j(t_k-\delta)}}{\sqrt{\la - E}} \right)+
\]
\beq\label{al2}
+ \sqrt{\la - E}\int_{t_k-\delta}^{t_k}
\frac{\pa}{\pa \la} \breve G_j(\xi_j(s,\la),s)d s,\quad\si_j\in\{-1,\,1\},
\eeq
where $\xi(\la,s) \in \mathcal{F}(E)$ such that $\mu_j(t_k-\delta)
\leq\xi(\la,s)\leq\la$ for
$t_k-\delta\leq s\leq t_k$. Furthermore, note that the function
\[
\breve G(\xi,s)=\frac{Y^{1/2}(\xi)}{\sqrt{\xi - E}\prod_{l\neq j}
(\xi - \mu_l)}
\]
is smooth with respect to $\xi$ in the domain $\mu_j(t_k-\delta)
\leq\xi\leq\la$ and
takes pure imaginary values there. Namely,
\[
\begin{array}{lllc} Y^{1/2}(\xi)\in\I\R,&\sqrt{\xi-E}
\in\R&\mbox{for}  & E\leq\xi\leq\la,\\
Y^{1/2}(\xi)\in\R, &\sqrt{\xi-E}\in\I\R& \mbox{for} &
\mu_j(t_k-\delta)\leq\xi\leq E.\end{array}
\]
Thus,
\beq\label{al4}
\frac{\pa^s\breve
G(\xi,s)}{\pa\xi^s}\in\I\R\quad \mbox{for}\quad
\mu_j(t_k-\delta)\leq\xi\leq\la,\quad s=0,1,\dots.
\eeq
The same considerations show
\beq\label{al6}
\sqrt{\la - E}=\dot\la
f_2(\la)\quad\mbox{where}\quad f_2(\la)\in\mathcal{F}(E),\,f(E)\neq 0.
\eeq
Combining this with \eqref{al4} we obtain
\[
\sqrt{\la - E}\int_{t_k-\delta}^{t_k}\frac{\pa}{\pa \la}\breve
 G_j(\xi_j(s,\la),s)ds=
\I\dot\la \,f_3(\la),\quad f_3(\la)\in\mathcal{F}(E).
\]
Thus
\beq\label{al5}\frac{d}{d\theta} \left(\sqrt{\la -
E}\int_{t_k-\delta}^{t_k}\breve G_j^\prime(\xi_j(s,\la),s)d s\right)
= \I f_4(\la), \quad f_4(\la)\in\mathcal{F}(E).
\eeq
Using
\eqref{al6} one can also represent the argument of $\arctan$  in the
first summand of \eqref{al2} as $\frac{f_5(\la)}{\dot\la}$, where
$f_5(\la)\in\mathcal{F}(E)$ and $f_5(E)\neq 0$. Therefore,
\beq\label{al7}
-\si_j\I \frac{d}{d\theta}\left(\arctan
\frac{\sqrt{E-\mu_j(t_k-\delta)}}{\sqrt{\la - E}}
\right) = -\si_j\I\frac{f_5^\prime(\dot\la)^2
-\ddot\la\,f_5}{(\dot\la)^2 +f_5^2 }\in\I \mathcal{F}(E).
\eeq
The same is valid for the interval $(t_k, t_k+\delta)$. Combining
\eqref{al1}, \eqref{al5}, and \eqref{al7} we obtain \eqref{al8}.
These considerations also show that the restriction \eqref{al10} is
unessential.
\end{proof}

Our next goal is to prove formula \eqref{justification}. Since for
 any solution of the equation $L_v(t) u=\la u$ the equality
$\mathcal{A}_vu=u_t - P_v(t)u$ is valid, it suffices to prove the following

\begin{lemma}\label{justprove}
Let $K_\pm(x,y,t)$ be the solutions of the GLM equations \eqref{ME1}
with the kernels \eqref{6.2}, corresponding to the
scattering data \eqref{refl}--\eqref{norm}. Let the functions
$\phi_\pm(\la,x,t)$ be defined by \eqref{Jost t} and
let $q(x,t)$ be defined by \eqref{5.111}. Then $\phi_\pm(\la,x,t)$ satisfy
\beq\label{just8}
\Big(\frac{\pa }{\pa t} - P_q(t) \Big) \phi_\pm(\la,x,t)=\beta_\pm(\la,t)
 \phi_\pm(\la,x,t),
\eeq
where $\beta_\pm(\la,t)$ is defined by \eqref{just3}.
\end{lemma}

\begin{proof}
As before we prove this lemma only for the $+$ case.
To simplify notations, set
$P=P_q(t)$, $P_0=P_+(t)$, $\phi=\phi_+(\la,x,t)$,
 $\psi=\psi_+(\la,x,t)$, $p=p_+$,
\[
(\mathcal{K} f)(x,t)=\int_x^{+\infty} K_+(x,y,t)f(y,t)dy
\]
\beq\label{d4}
(\dot{\mathcal{K}} f)(x,t)=\int_x^{+\infty} \frac{\pa}{\pa t}
K_+(x,y,t)f(y,t)dy,
\eeq
and denote by a dot the derivative with respect to $t$ and by
a prime the derivative with respect to spatial variables.
Moreover, we will omit the variable $t$ whenever it is possible
 and use the notations
\[
D_{x^l y^m}(x):=\left(\frac{\pa^l}{\pa x^l} + \frac{\pa^m} {\pa
y^m}\right) D(x,y)|_{y=x}, D_{x^0 y^0}(x)= D(x).
\]
Since $\dot\psi - P_0\psi=\beta\psi$, then
\beq\label{derivat}
\dot\phi - P\phi=\beta\phi + (P_0 - P)\psi + \dot{\mathcal{K}}
\psi +\mathcal{K} P_0\psi - P\mathcal{K} \psi.
\eeq
Differentiating  the last term and integrating by parts gives
\begin{align}\nn
(P\mathcal K \psi)(x)= &\left\{-2(q^\prime(x) - p^\prime(x))
+4 K_{xy}(x) + 8 K_{x^2}(x) - 6 q(x)K(x)\right\}\psi(x)\\\nn
& -\left\{4(q(x) - p(x)) -  4 K_x(x)\right\}\psi^\prime(x) +
 4K(x)\psi^{\prime\prime}(x) +\\\label{d1}
&+\int_x^\infty\left(-4 K_{x^3}(x,y) + 6 q(x)K_x(x,y) +
3q^\prime(x)K(x,y)\right)\psi(y)dy,
\end{align}
and
\begin{align}\nn
\left(\mathcal{K} P_0 \psi\right)(x) = & \left(4 K_{y^2}(x) -
6 K(x)p(x)\right)\psi(x) - 4 K_y(x)\psi^\prime(x) +
4K(x)\psi^{\prime\prime}(x)\\\label{d2}
& + \int_x^\infty \left(4 K_{y^3}(x,y) - 6 K_y(x,y)p(y) -
3 K(x,y)p^\prime(y)\right)\psi(y)dy.
\end{align}
Besides,
\beq\label{d3}
(P - P_0)\psi(x)=6(q(x) - p(x))\psi^\prime(x) + 3(q^\prime(x)
 - p^\prime(x))\psi(x).
\eeq
Combining \eqref{d4}--\eqref{d3} and taking into account the
formula (cf.\ \cite{F1})
\beq\label{fir1}
-K_{xx}(x,y) + q(x) K(x,y) = -K_{yy}(x,y) + p(y) K(x,y),
\eeq
where we put $x=y$, we arrive at the representation
\beq\label{derivat2}
(\dot\phi - P\phi-\beta\phi)(x) =A(x)\psi(x) + B(x)
\psi^\prime(x) +\int_x^\infty (\tau^{xy}K(x,y))\psi(y)dy=0,
\eeq
where
\begin{align*}
A(x) &= p^\prime(x) - q^\prime(x) - 2 K_{x^2}(x) - 4K_{xy}(x)
 - 2K_{y^2}(x),\\
B(x) &= 2(p(x) - q(x)) - 4(K_x(x) + K_y(x)),
\end{align*}
and
\beq\label{tauxy}
\tau^{xy}:=\frac{\pa}{\pa t} + \tau_q^x +\tau_p^y,\quad\tau_q^x:=
4\frac{\pa^3}{\pa x^3} -6 q(x)\frac{\pa}{\pa x} - 3 q^\prime(x).
\eeq
But according to \eqref{5.111}
\beq\label{fir2}
p(x) - q(x)=2K_x(x) + 2K_y(x), \quad p^\prime(x) - q^\prime(x)=
2K_{x^2}(x) + 4K_{xy}(x) +2 K_{y^2}(x),
\eeq
and therefore, $A(x)=B(x)=0$. Thus, to prove \eqref{just8} one
 has to check, that
\begin{align}\nn
D(x,y):= & \tau^{xy}K(x,y)= K_t(x,y) +4 K_{y^3}(x,y) + 4 K_{x^3}
(x,y)- 6q(x)K_x(x,y)\\\label{Dfin}
& -6p(y)K_y(x,y) - 3q^\prime(x)K(x,y) - 3p^\prime(y)K(x,y)\equiv 0.
\end{align}
To this end, let us derive an equation for the function $F=F_+(x,y,t)$,
 defined by formula \eqref{6.2}.
This function can be represented (see \eqref{6.2hat}) as
\[
F(x,y,t)=\int_\mathbb{R}\hat\psi(\la,x,t)\hat\psi(\la,y,t)d\rho(\la),
\]
where the measure
\begin{align*}
d\rho(\la)= &\Big(\frac{1}{\pi \I}R_+(\la,0)g_+(\la,0)\chi_{\si_+^u}
(\la) +\frac{1}{2\pi \I}\vert T_-(\la,0) \vert^2 g_-(\la,0)
\chi_{\si_-^{(1)}}(\la)\\
& +\sum_{k}(\gamma_k^+)^2(0)\delta(\la - \la_k)\delta_+
(\la_k,0)^2 \Big)d\la
\end{align*}
does not depend on $t$.
Using \eqref{LP2} we conclude, that
\beq\label{tauf}
\tau_0^{xy}F(x,y)=0,\quad \tau_0^{xy}=\frac{\pa}{\pa t} +
\tau_p^x  + \tau_p^y.
\eeq
Now set $V(x)=q(x) - p(x)$ and apply the operator $\tau^{xy}$
 to the GLM equation \eqref{ME1}.
Taking into account \eqref{Dfin}, \eqref{tauf} and the equality
\[
\tau^{xy} - \tau_0^{xy}=-6V(x)\frac{\pa}{\pa x} - 3V^\prime(x)
\]
we obtain
\begin{align*}
D(x,y) = & \int_x^\infty\left\{
K(x,s)\tau_p^s\left[F(s,y)\right]- K_t(x,s)F(s,y)\right\}ds\\
& -\tau_q^x\left[\int_x^\infty K(x,s)F(s,y)ds\right]+6V(x)F_x(x,y)
 + 3V^\prime(x)F(x,y),
\end{align*}
or
\beq\label{Deq1}
D(x,y) + \int_x^\infty D(x,s)F(s,y)ds=r(x,y),
\eeq
where
\beq\label{req}
r(x,y)=\int_x^\infty \left\{\tau_p^s\left[K(x,s)\right]F(s,y)
+K(x,s)\tau_p^s\left[F(s,y)\right]\right\}ds +
\eeq
\[
+\int_x^\infty \tau_q^x\left[K(x,s)\right]F(s,y)ds -
\tau_q^x\left[\int_x^\infty K(x,s)F(s,y)ds\right]+
\]
\[
+6V(x)F_x(x,y) + 3V^\prime(x)F(x,y).
\]
It is proved in \cite{BET}, that the equation $D(x,y) +
\int_x^\infty D(x,s)F(s,y)ds=0$, where $x$ plays the role of a
parameter, has only the trivial solution in the space
$L^1(x,\infty)$. Since the function $D(x,\cdot)$ evidently belongs
to this space, then to prove \eqref{Dfin} it is sufficient to prove
that $r(x,y)=0$.

Taking into account, that $V(x)=-2\frac{d}{dx}K(x,x)$,
direct computations imply
\beq\label{tauxk}
\int_x^\infty \tau_q^x\left[K(x,s)\right]F(s,y)ds -
\tau_q^x\left[\int_x^\infty K(x,s)F(s,y)ds\right] +
\eeq
\[
+6V(x)F_x(x,y) + 3V^\prime(x)F(x,y)=4 K(x,x)F_{x^2}(x,y) +
\]
\[
+4K_x(x,x)F_x(x,y)+8K_{x^2}(x,x)F(x,y) +4K_{xy}(x,x)F(x,y) +
\]
\[
+V^\prime(x)F(x,y) +2V(x)F_x(x,y) - 6 q(x)K(x,x)F(x,y).
\]
From the other side, integration by parts gives
\beq\label{tausf}
\int_x^\infty \left\{\tau_p^s\left[K(x,s)\right]F(s,y) +
K(x,s)\tau_p^s\left[F(s,y)\right]\right\}ds =
\eeq
\[
=-4\left\{K_{s^2}(x,s)F(x,y) +K(x,x)F_{s^2}(s,y)-K_s(x,s)
F_s(s,y)\right\}|_{s=x}+
\]
\[
+6p(x)K(x,x)F(x,y).
\]
Substituting last to formulas to \eqref{req} gives
\[
r(x,y)=F_x(x,y)(4K_x(x,x) + 4K_y(x,x) +2V(x)) +
\]
\[
+ F(x,y)\left( -6V(x)K(x,x) +8K_{x^2}(x,x) +4K_{xy}(x,x) -
 4K_{y^2}(x,x) +V^\prime(x)\right).
\]
Taking into account \eqref{fir2} we obtain
\[
r(x,y)=F(x,y)\left(-6V(x)K(x,x)+6K_{x^2}(x,x) -
6K_{y^2}(x,x)\right),
\]
and \eqref{fir1} implies $r(x,y)=0$.
\end{proof}

\noindent{\bf Acknowledgments.} We are very grateful to
V.A. Marchenko and E.Ya. Khruslov for helpful discussions.
I.E. gratefully acknowledge the extraordinary hospitality
of the Faculty of Mathematics
at the University of Vienna during extended stays 2008--2009,
 where parts of this paper were written.

\end{document}